\documentclass[useAMS,usenatbib]{mnras}
\pdfoutput=1
\usepackage{deluxetable}
\usepackage[]{natbib}
\usepackage{amsmath} 
\usepackage{amssymb} 
\usepackage{gensymb}
\usepackage{graphicx} 
\usepackage{indentfirst} 
\usepackage{pdflscape} 
\usepackage{placeins} 
\usepackage{nicefrac} 
\usepackage{afterpage} 
\usepackage{rotating} 
\usepackage{booktabs} 
\usepackage{rotating} 
\usepackage{color} 
\usepackage{longtable}
\usepackage{amsmath}
\usepackage{csquotes}
\usepackage{threeparttable}
\usepackage{enumitem}
\usepackage{dblfloatfix}

\usepackage{mathrsfs}

\newcommand{\HI}{\mbox{{H\,{\scriptsize I}}}}

\newcommand{\MgII}{{\mbox{Mg\,{\scriptsize II}\ }}}
\newcommand{\NII}{{\mbox{N\,{\scriptsize II}}}}
\newcommand{\CIV}{{\mbox{C\,{\scriptsize IV}\ }}}

\newcommand{\OII}{{\mbox{[O\,{\scriptsize II}]}}}
\newcommand{\OIII}{{\mbox{[O\,{\scriptsize III}]}}}
\newcommand{\OVI}{{\mbox{O\,{\scriptsize VI}}}}
\newcommand{\Halpha}{{\mbox{H$\alpha$}}}
\newcommand{\Hbeta}{{\mbox{H$\beta$}}}

\newcommand{\MAGMAGMAG}{M3}
\newcommand{\mstar}{M_{\rm star}}
\newcommand{\msun}{{\rm M}_\odot}

\newcommand{\apg}{\gtrsim}
\newcommand{\apl}{\lesssim}

\newcommand{\ewr}{\mbox{$W_r(2796)$}}
\newcommand{\kms}{\,${\rm{km\,s}^{-1}}$}

\title[Magellan MagE \MgII (M3) Halo Project]{A Complete Census of Circumgalactic \MgII at Redshift z $\lesssim$ 0.5\thanks{Based on data gathered with the 6.5m Magellan Telescopes located at Las Campanas Observatory.}}
\author[Huang et al.]{Yun-Hsin Huang$^{1}$\thanks{E-mail: 
yunhsin@email.arizona.edu}, Hsiao-Wen Chen$^{2}$, Stephen A.\ Shectman$^{3}$, 
Sean D.\ Johnson$^{4}$,
 \newauthor Fakhri S.\ Zahedy$^{3}$, Jennifer E.\ Helsby$^{5}$, Jean-Ren\'{e} Gauthier$^{6}$, Ian B.\ Thompson$^{3}$\\ \\ 
$^{1}$Steward Observatory, University of Arizona, Tucson, AZ 85721, USA \\
$^{2}$Department of Astronomy \& Astrophysics, and Kavli Institute for 
Cosmological Physics, The University of Chicago, Chicago, IL 60637, USA\\
$^{3}$The Observatories of the Carnegie Institution for Science, 813 Santa Barbara Street, Pasadena, CA 91101, USA \\
$^{4}$Department of Astronomy, University of Michigan, 1085 S. University Ave, Ann Arbor, MI 48109, USA \\
$^{5}$Freedom of the Press Foundation, 601 Van Ness Ave, San Francisco, CA 94102 \\
$^{6}$Oracle Corporation, Redwood Shores, CA 94065 \\
}
\begin{document}

\date{\today}

\pagerange{\pageref{firstpage}--\pageref{lastpage}} \pubyear{2002}

\maketitle

\label{firstpage}

\begin{abstract}
This paper presents a survey of \MgII absorbing gas in the vicinity of
380 random galaxies, using 156 background quasi-stellar objects (QSOs)
as absorption-line probes.  The sample comprises 211 isolated (73
quiescent and 138 star-forming galaxies) and 43 non-isolated galaxies
with sensitive constraints for both \MgII absorption and \Halpha\
emission.  The projected distances span a range from $d=9$ to 497 kpc,
redshifts of the galaxies range from $z=0.10$ to 0.48, and rest-frame
absolute B-band magnitudes range from $M_{\rm B}=-16.7$ to $-22.8$.
Our analysis shows that the rest-frame equivalent width of
{Mg\,{\scriptsize II}}, \ewr, depends on halo radius ($R_h$),
$B$-band luminosity($L_{\rm B}$) and stellar mass ($M_{\rm star}$) of
the host galaxies, and declines steeply with increasing $d$ for
isolated, star-forming galaxies.  At the same time, \ewr\ exhibits no
clear trend for either isolated, quiescent galaxies or non-isolated
galaxies.  In addition, the covering fraction of \MgII absorbing gas
$\langle \kappa \rangle$ is high with $\langle \kappa \rangle\gtrsim
60$\% at $<40$ kpc for isolated galaxies and declines rapidly to
$\langle \kappa \rangle\approx 0$ at $d\gtrsim100$ kpc.  Within
the gaseous radius, the incidence of \MgII gas depends sensitively on both
$M_{\rm star}$ and the specific star formation rate inferred from
\Halpha.  Different from what is known for massive quiescent halos,
the observed velocity dispersion of \MgII absorbing gas around
star-forming galaxies is consistent with expectations from virial
motion, which constrains individual clump mass to $m_{\rm cl} \gtrsim
10^5 \,\rm M_\odot$ and cool gas accretion rate of $\sim 0.7-2
\,M_\odot\,\rm yr^{-1}$.  Finally, we find no strong azimuthal
dependence of \MgII absorption for either star-forming or quiescent
galaxies.  Our results demonstrate that multiple parameters affect the
properties of gaseous halos around galaxies and highlight the need of
a homogeneous, absorption-blind sample for establishing a holistic
description of chemically-enriched gas in the circumgalactic space.
%
\end{abstract}


\section{Introduction}
Over the past decades, extensive progress has been made to understand
the impact of the baryon cycle on galaxy formation and evolution, 
with particular focus on gas reservoirs such as the circumgalactic medium (CGM).  
Located in the space between galaxies and the intergalactic medium (IGM), 
the CGM contains critical information on gas accretion and outflows, 
processes that drive the evolution of galaxies \citep[see][for recent reviews]{Chen:2017a,Tumlinson:2017}. Thus, the CGM provides an excellent laboratory for understanding the physical processes that drive the formation and evolution of galaxies.

Absorption-line spectroscopy of background quasars has provided 
a unique probe of the low-density CGM,
which is otherwise too diffuse to be detected in emission beyond the local Universe. 
Over the last decade, statistically significant samples of galaxies 
at $z\approx0-2$ have been assembled using a combination of 
space- and ground-based telescopes. 
The Cosmic Origins Spectrograph (COS) on the 
{\it Hubble Space Telescope} ({\it HST}) has enabled
studies of a rich suite of absorption lines including the HI Lyman series \cite[e.g.,][]{Chen:1998,Tripp:1998,Rudie:2013,Tumlinson:2013,Werk:2014,Liang:2014,Borthakur:2016}, the \OVI\ doublet \cite[e.g.,][]{Chen:2009,Prochaska:2011,Tumlinson:2011,Johnson:2015}
and the \CIV doublet \cite[e.g.,][]{Borthakur:2013,Bordoloi:2014,Liang:2014}.

From the ground, at $z\lesssim2$ the majority of 
studies have focused on the measurement
of the \MgII $\rm \lambda \lambda\,2796,2803$ doublets
due to their strength and visibility in the optical range.
This transition is thought to arise primarily in 
photoionized gas of temperature $T\sim 10^4\,\rm K$
 \citep{Bergeron:1986,Charlton:2003}
and high neutral hydrogen column density clouds of 
$N(\HI)\approx10^{18}-10^{22}\,\rm cm^{-2}$ \citep{Rao:2006} {for \ewr\ $>0.3$\AA\ absorbers}.
Many investigations have been carried out to characterize
the statistical properties of \MgII absorbers, 
including the frequency distribution function, 
redshift evolution of the absorber number density 
and kinematic signatures  \citep[e.g.][]{Lanzetta:1987,Petitjean:1990,Charlton:1998,Churchill:2000,Churchill:2003,Nestor:2005}.  
Studies also show that \MgII absorbing gas probes the
underlying gas kinematics around galaxies \citep[e.g.,outflow and inflow gas.][]{Weiner:2009,Kacprzak:2012,Rubin:2014,Ho:2017,Ho:2020}.  

To have a comprehensive understanding of the 
baryonic structures around galaxies, significant 
progress has been made in constructing samples 
of galaxy-\MgII absorber pairs to
understand the correlation between
cool, enriched gas and galaxy properties.
On the one hand, some galaxy-\MgII pair associations begin
with quasar spectra and then search for 
nearby galaxies responsible for the \MgII
absorption \citep[e.g.][]{Kacprzak:2011}.  
Such studies commonly target galaxies already known to have
\MgII absorption in the spectra of background quasars
and therefore may result in biased galaxy populations.
On the other hand, studies have investigated relationships
between galaxies and their surrounding gas using unbiased 
samples, where the galaxy-QSO pairs are chosen 
without any prior knowledge of the presence or absence 
of absorbing gas, allowing detailed studies of the CGM 
as a function of galaxy properties (including stellar mass, 
star formation rate and color) and environment
\citep[e.g.][]{Chen:2010,Johnson:2015b,Huang:2016,Zahedy:2016,Lan:2018,Martin:2019}.

As the \MgII doublet features start to be observable 
in the optical wavelengths at $z\sim0.4$, 
this transition has not been studied as extensively at $z\lesssim0.4$.
Here we make use of the UV sensitive spectrograph, 
the Magellan Echellette Spectrograph \citep[MagE;][]{Marshall:2008}, 
to perform searches for \MgII absorbers at redshifts as low as $z\sim0.1$.
Building upon the SDSS database, we conduct the Magellan 
MagE \MgII (\MAGMAGMAG) Halo Project in the 
spectra of background QSOs at $z\lesssim0.4$.
The main goal of the \MAGMAGMAG\ Halo Project is to 
establish an unbiased, statistically significant sample of 
$z\lesssim0.4$ \MgII absorbers to constrain the incidence, 
strength and extent of \MgII absorbing gas around 
galaxies of different properties.

The first-year results of the \MAGMAGMAG\ Halo Project 
is reported in \cite{Chen:2010} (hereafter C10).
With a spectroscopic sample of 94 galaxies at a median 
redshift of $\langle z \rangle=0.24$
and projected distance $d \lesssim170\,\rm kpc$, 
\cite{Chen:2010,Chen:2010b}
investigated the possible correlations between the 
incidence and extent of \MgII absorbers
and galaxy properties.  We found the rest-frame 
equivalent width of \MgII (\ewr) 
declines steeply with increasing $d$ from the galaxies.  
Moreover, the extent of
\MgII gaseous halos scales strongly with the galaxy 
$B-$band luminosity and galaxy stellar
masses, with slight dependence on specific star 
formation rate (sSFR) 
and no dependence on galaxy $B_{AB}-R_{AB}$ color.
The first-year results clearly demonstrate that \MgII 
absorbing gas is strongly connected to the physical properties of host galaxies. 
Using the full sample of the \MAGMAGMAG\ Halo Project, 
we will show that we are able to observe a clear difference 
in surrounding gas properties between star-forming and quiescent galaxies.

The paper is organized as follows.
In Section \ref{section:observation}, we describe the 
experimental design of the \MAGMAGMAG\ Halo Project, 
and the spectroscopic observations and data reduction of 
the photometrically selected galaxies and spectroscopically 
confirmed quasars in the SDSS archive.  
We present the catalogs of galaxies and \MgII absorbers in 
Section \ref{section: catalogs}.
In Section \ref{section:analysis}, we describe our likelihood 
analysis and characterize the
correlation between \MgII absorption strength and galaxy properties.
In Section \ref{section:discussion}, we discuss the covering fraction, 
the kinematics and 
the azimuthal dependence of \MgII absorbing gas.
We discuss the difference between different types of galaxies, 
the difference between isolated and non-isolated galaxies, and compare
our results with previous studies. We present
a summary of our findings in Section \ref{section:summary}.
We adopt the standard $\Lambda$ cosmology, $\Omega_M$ = 0.3 and
$\Omega_\Lambda$=0.7 with a Hubble constant $H_{\rm 0} = 70\rm \,km\,s^{-1}\,Mpc^{-1}$.

\begin{table*}
	\caption{Summary of Faint Galaxy Spectroscopy}
	\label{table:galSummary}
	\begin{tabular}{cccccccc}
	\hline \hline
ID & RA(J2000) & Dec(J2000) & $z_{\rm phot}$ & $r$ & Instrument & Exptime & UT Date \\
	\hline 
SDSSJ000548.29$-$084757.25 & 00:05:48.29 & $-$08:47:57.25 & $0.25 \pm 0.04$ & 19.2 & MagE & $600        $ & 2009 Oct  19   \\  
SDSSJ000548.59$-$084801.16 & 00:05:48.59 & $-$08:48:01.14 & $0.22 \pm 0.07$ & 19.4 & MagE & $300+600    $ & 2009 Oct  19   \\  
SDSSJ001335.12+141439.54 & 00:13:35.12 & +14:14:39.55 & $0.26 \pm 0.11$ & 20.9 & DIS & $3\times1800     $ & 2008 Dec  22   \\  
SDSSJ001336.14+141428.04 & 00:13:36.14 & +14:14:28.01 & $0.25 \pm 0.08$ & 19.5 & DIS & $1800+1200  $ & 2008 Dec  22   \\  
SDSSJ003009.52+011445.25 & 00:30:09.52 & +01:14:45.25 & $0.15 \pm 0.02$ & 17.7 & SDSS&...&... \\  
SDSSJ003009.90+011343.95 & 00:30:09.90 & +01:13:43.95 & $0.53 \pm 0.04$ & 21.0 & SDSS&...&... \\  
SDSSJ003010.18+011219.92 & 00:30:10.18 & +01:12:19.92 & $0.17 \pm 0.03$ & 18.5 & SDSS&...&... \\  
SDSSJ003012.84+011131.36 & 00:30:12.84 & +01:11:31.36 & $0.08 \pm 0.02$ & 17.0 & SDSS&...&... \\  
SDSSJ003014.16+011359.19 & 00:30:14.17 & +01:13:59.14 & $0.40 \pm 0.11$ & 20.8 & DIS & $1800       $ & 2009 Nov  9    \\  
SDSSJ003016.41+011406.90 & 00:30:16.41 & +01:14:06.90 & $0.74 \pm 0.05$ & 21.2 & SDSS&...&... \\  
	\hline
	\multicolumn{8}{l}{The full table is available in the on-line version of the paper.
}
	\end{tabular}
\end{table*}

\section{OBSERVATIONS}
\label{section:observation}
\subsection{Experiment Design}
\label{section:exp}
To investigate the correlation between galaxy properties and \MgII absorbing gas
at small projected distances,
we need to obtain spectroscopic data of both galaxies and QSO absorbers
along common sightlines.
We utilize the Magellan Echellette Spectrograph
\citep[MagE;][]{Marshall:2008} on the Magellan Clay Telescope
to conduct a survey of \MgII absorbers at $z<0.4$.
The high UV throughput of MagE from $\lambda=3100\,$\AA\ enables
searches of \MgII absorbers at redshift as low as 0.11.
We refer the reader to C10 for a detailed description of the survey design.
Briefly, the QSO and galaxy pairs are selected from the SDSS DR6 catalogs \citep{Adelman:2008}.
To maximize the efficiency of searching \MgII absorbers, 
we consider galaxies at photometric redshifts of $z_{\rm phot}\leq0.4$
that have background QSOs in close projected distance $d<\hat R_{\rm gas}$.
{$\hat R_{\rm gas}=130\,{\rm kpc}$} is the distinct boundary
found by \cite{Chen:2008}
using 23 galaxy-QSO pairs at intermediate redshifts of $z\sim0.4$, 
beyond which no \MgII absorbers are found.
Note that although we pre-select galaxy-QSO pairs with $d<\hat R_{\rm gas}$ for
the spectroscopic followup survey, 
we also search the public SDSS DR14 sample \citep{Abolfathi:2018}
to include galaxies with spectroscopic redshifts at $z\leq0.4$ around our observed
QSOs to study the gaseous halo beyond $\hat R_{\rm gas}$.
{Therefore, the properties of \MgII absorbing gas around galaxies
are studied both within and beyond
 $\hat R_{\rm gas}$ in our survey.}
 {We reiterate that the galaxy-QSO pairs are chosen without 
 any prior knowledge of the presence or absence of absorbing gas. }
In the following sections, we describe the galaxy spectroscopic 
sample either obtained from
our own observations or SDSS DR14 archive 
and the observations of background quasars.

\subsection{Galaxy Spectroscopy}
To establish a physical connection between galaxies 
and \MgII absorbing systems 
along nearby QSO sightlines, it is essential to have medium 
to high resolution spectra to obtain
precise and accurate redshift measurements of these galaxies.
We have obtained optical spectra of 218 galaxies that 
satisfy the criteria described above
using the MagE Spectrograph \citep{Marshall:2008} at the
Las Campanas Observatory and the Double Imaging
Spectrograph \citep[DIS;][]{Lupton:1995}
on the 3.5 m telescope at the Apache Point Observatory.
Details about the spectroscopic observation setups and data
reduction are presented in C10.  
In summary, we acquired 120 long-slit galaxy spectra
using DIS over the period from 2008 August through 2010 September,
and echellette spectra of 98 galaxies
using MagE from 2008 August to 2011 March.
The spectra obtained using DIS and MagE
have intermediate resolution of FWHM $\apl 500$ \kms and $\approx$ 150 \kms
 in the wavelength range between $\lambda\,\approx$ 4000\AA\ and $1\,\rm \mu m$.
 We reduced the DIS spectroscopic data using standard long-slit spectral reduction
procedures, and the MagE spectra using the software developed by G. Becker
with a slight modification to work with binned spectral frames.
 The redshifts of these galaxies were determined 
using a cross-correlation
 analysis with a linear combination of SDSS galaxy eigen spectra  The
 typical redshift uncertainty is $\Delta z \sim 0.0003$ and 0.0001 for
 galaxy spectra taken using DIS and MagE.
 
 We include 17 additional SDSS DR14 galaxies which already have reliable spectroscopic redshifts in the SDSS archive with projected distance $d<\hat R_{\rm gas}$ in our galaxy sample.  We have also extended our search 
 to $d<500\,\rm kpc$ in the SDSS archive and located 145 SDSS galaxies with
 accurate spectroscopic redshifts available.  
 Combined with our own observations, we have a total of 380 galaxies at $d<500\, \rm kpc$ in our final galaxy sample for searches of \MgII absorbers.
A journal of the observations of the full galaxy 
sample is presented in Table \ref{table:galSummary}.
 
 \begin{table*}
\caption{Summary of the MagE Spectroscopic Observations of SDSS QSOs} 
\label{table:QSOSummary}
	\begin{tabular}{ccccccc}
	\hline \hline
	ID & RA(J2000) & Dec(J2000) & $z_{\rm QSO}$ & $u'$ & Exptime & UT Date \\
	\hline 
SDSSJ000548.24$-$084808.44 & 00:05:48.24 & $-$08:48:08.44 & 1.19 & 17.96 & $2\times1200$ & 2009 Oct 19 \\  
SDSSJ001335.75+141424.07 & 00:13:35.75 & +14:14:24.07 & 1.54 & 19.32 & $1800+1300$ & 2009 Oct 20 \\  
SDSSJ003013.91+011405.14 & 00:30:13.91 & +01:14:05.14 & 1.46 & 18.12 & $2\times1800$ & 2010 Jul 13 \\  
SDSSJ003340.21$-$005525.53 & 00:33:40.21 & $-$00:55:25.53 & 0.94 & 17.99 & $2\times900$ & 2008 Sep 23 \\  
SDSSJ003407.35$-$085452.12 & 00:34:07.35 & $-$08:54:52.12 & 1.31 & 18.59 & $2\times1200$ & 2008 Sep 24 \\  
SDSSJ003413.04$-$010026.86 & 00:34:13.04 & $-$01:00:26.86 & 1.29 & 17.33 & $2\times600$ & 2008 Sep 23 \\  
SDSSJ010135.84$-$005009.08 & 01:01:35.84 & $-$00:50:09.08 & 1.01 & 19.31 & $2\times1800$ & 2008 Sep 24 \\  
SDSSJ010156.32$-$084401.74 & 01:01:56.32 & $-$08:44:01.74 & 0.98 & 18.29 & $2\times1800$ & 2008 Sep 25 \\  
SDSSJ010205.89+001156.99 & 01:02:05.89 & +00:11:56.99 & 0.72 & 17.59 & $1500+900$ & 2009 Oct 18 \\  
SDSSJ010352.47+003739.79 & 01:03:52.47 & +00:37:39.79 & 0.70 & 18.36 & $3\times1200$ & 2008 Sep 25 \\  
	\hline
	\multicolumn{7}{l}{The full table is available in the on-line version of the paper.}
	\end{tabular}
\end{table*}
 
\subsection{Echellette Spectra of QSOs}
Echellete spectroscopic observations of 156 QSOs were obtained
using the MagE spectrograph \citep{Marshall:2008} on the Magellan Clay telescope 
over the period from 2008 January through 2011 June.
The majority of QSOs were observed using a $1\arcsec$slit, 
yielding a typical spectral resolution of FWHM $\approx$ 70 \kms.  
All the QSO spectra were processed and reduced using the data
reduction software developed by G. Becker.
We refer the reader to C10 for details of observations and spectral 
reduction procedures. In brief, we first performed the wavelength and 
flux calibrations, and co-added the individual echellete orders to form 
a single contiguous spectrum across
the spectral range from $\lambda = 3050$ \AA\ to $\lambda=1\rm\, \mu m$.
The individual order-combined exposures were then continuum 
normalized and finally were stacked to form the final reduced spectrum. 
{Ths S/N per resolution element is $\apg$ 10 at wavelengths greater than $\sim 3200\,$\AA ($z\sim$ 0.14), leading to a 2$\sigma$ upper limit of \ewr\ $\lesssim 0.13 $\AA.
For galaxies at lower redshifts, the error blows up and 
the S/N per resolution element can be as low as $\sim 3$.}
We present a journal of the spectroscopic observations of the QSOs in Table \ref{table:QSOSummary}.

\section{THE GALAXY AND \MgII ABSORBER CATALOGS}
\label{section: catalogs}
\subsection{Galaxy properties}
\label{section:galproperties}
We have constructed a full sample of 380 galaxies with 
robust redshift measurements in the vicinity of 156 distant 
background QSO sightlines.
Among these galaxies, 103 galaxies are found to have
at least one spectroscopic neighbor at projected distance $d\leq500\,\rm kpc$ 
and radial velocity difference $\Delta v $ smaller than 1000 \kms.
The presence of close neighbors imply that these galaxies are likely
 to reside in a group environment, where
the interactions of group members may change the correlation 
between galaxy properties and their gaseous halos. 
We also performed a literature search and identified 9 galaxies 
that are either previously known merging systems, 
galaxy groups or clusters 
\citep[e.g.,][]{Koester:2007,Hao:2010,Smith:2012,Johnson:2014}.
To avoid the confusion of associating \MgII absorbers with 
host galaxies, we classify these galaxies as \enquote{non-isolated} 
galaxies and discuss them separately.
The criteria yielded a sample of 277 \enquote{isolated} galaxies and
103 \enquote{non-isolated} galaxies.

\begin{figure}
	\centering
	\includegraphics[scale=0.44]{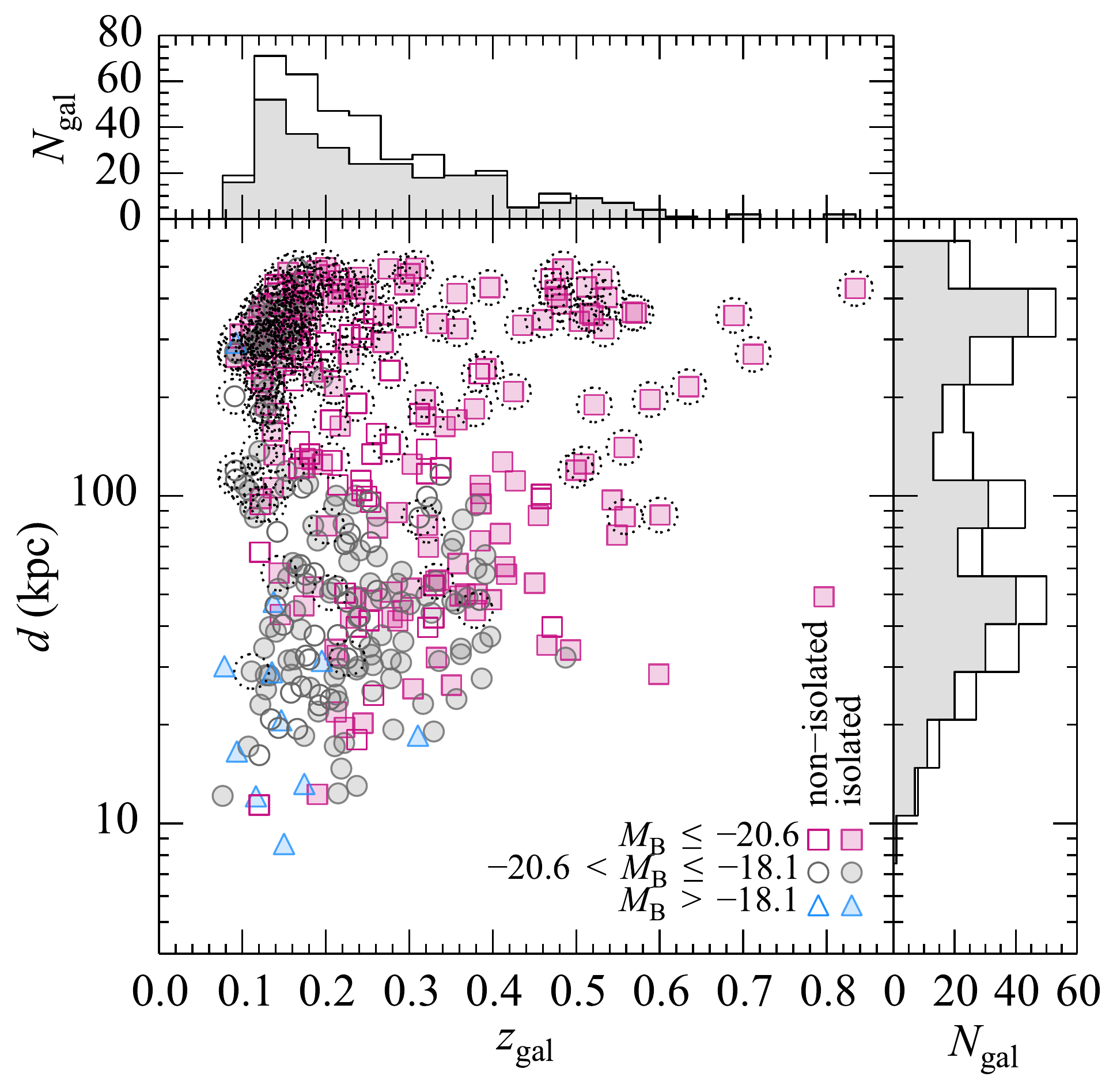}
	\caption{
	Distribution of projected distance $d$ versus
	redshift for 277 isolated galaxies (solid symbols)
	and 103 galaxies in the \enquote{non-isolated} sample (open symbols).
	We show luminous galaxies of $L_B>L_{B_*}$
	 in magenta squares, sub-L* galaxies of
	$L_B = (0.1-1)\,L_{B_*}$ in grey circles, and 
	low-luminosity galaxies of $L_B<0.1\,L_{B_*}$ in cyan triangles.
	Galaxies with spectroscopic redshifts from 
	SDSS are marked in dotted circles.
	The $top$ and $right$ panels show respectively
	 the redshift and projected distance histograms of 
	 the full sample (open histograms)
	 and the isolated galaxy sample (filled histograms).} 
	\label{figure:zgal_d}
\end{figure}

We first present the projected distance $d$ versus redshift 
distribution of the full galaxy sample in Figure \ref{figure:zgal_d}.  
The isolated and non-isolated galaxies are presented in 
solid and open symbols. The redshifts of the galaxies range 
from $z=0.08$ to $z=0.83$ with a median of $\left<z\right>=0.22$.
Using our own DIS and MagE observations, we measure 
redshifts of 218 galaxies and find that the SDSS photometric 
redshift measurements are accurate to within a median residual
of $\left<|(z_{\rm phot}-z_{\rm spec})/(1+z_{\rm spec})|\right>$ = 0.05 
with a rms scatter of $\delta z \equiv \Delta z /(1+z) = 0.07$.
We find that 91\,\% of galaxies (198 out of 218) have redshifts 
at $z=0.1-0.4$, showing that the target selection based on 
photometric redshifts is effective in identifying foreground
galaxies at our targeted redshift range. 

For our own observations, the projected distances of the 
QSOs range from $d\approx9-178$ kpc with a median of 
$\left<d\right>=48$ kpc.  
18\% (39 out of 218) galaxies have projected 
distance $d>\hat R_{\rm gas}$, and only one galaxy 
has $d>1.5\,\hat R_{\rm gas}$, showing that our QSO-galaxy 
pair selection effectively probes properties of gaseous 
halos close to the host galaxies. The SDSS galaxies occupy 
predominantly the regime of large projected distances, where 
40 galaxies have $d\leq1.5\,\hat R_{\rm gas}$ and the remaining 
122 galaxies extend out to $d=497\,\rm kpc$.
The full galaxy sample has $d\approx9-497$ with a median 
of $\left<d\right>=94$ kpc. In Figure  \ref{figure:zgal_d}, we also 
adopt various colors and symbols to
highlight different ranges of galaxy luminosity.
We estimate a $B$-band luminosity for each galaxy following 
the steps in \cite{Chen:2010}, and find that the full galaxy sample 
spans a wide range of $B$-band luminosity.
The sample has 200 luminous galaxies ($L_{\rm B}>L_{\rm B_*}$) 
spanning a range in their projected distance to a QSO sightline 
from $d=11\,\rm kpc$ to $d=497\,\rm kpc$,
and 169 sub-$L_*$ galaxies ($L_{\rm B}= 0.1-1 \, L_{\rm B_*}$) 
covering a range from $d=12\,\rm kpc$ to $d=494\,\rm kpc$.  
The sample also includes 11 low-luminosity dwarf galaxies with
$L_{\rm B} < 0.1\,L_{\rm B_*}$ from $d=9\,\rm kpc$ to  $d<293\,\rm kpc$.

While $B$-band luminosity is known to scale with halo mass \citep[e.g.,][]{Yang:2005,Tinker:2007},
it is also found to correlate with \OII\ luminosity despite a large
scatter \citep{Zhu:2009}. To have a more robust tracer of dark 
matter halo mass, we also calculate the total stellar mass 
$M_{\rm star}$ \citep{More:2011} and \Halpha\ equivalent 
width as an indicator of star formation.
We estimate the stellar mass according to \citet[Equation 1 and 2]{Johnson:2015}.
Briefly, they use low-redshift galaxies of $z<0.055$ 
in the NASA-Sloan Atlas \citep[e.g.][]{Maller:2009} to derive 
the relation between stellar mass and 
rest-frame $g$- and $r$- band absolute magnitudes.
The relation can well reproduce the NASA-Sloan altas stellar
masses with a systematic error of less than 0.02 dex and 
a 1-$\sigma$ scatter of less than 0.15 dex over the entire
range of our galaxy absolute $r$-band magnitudes.
They also show that the systematic error induced by the 
redshift evolution of the mass-to-light ratio relation for 
$z=0.1-0.4$ galaxies is less than 0.1 dex.  
We estimate the rest-frame $g$- and $r$- band absolute 
magnitudes by interpolation of observed SDSS $g,r,i,z$ bands.  
For fainter galaxies with $\leq 5$-$\sigma$ detection in 
$g$-band absolute magnitude, we calculate the stellar mass 
using the relation between stellar mass and single rest-frame 
$r$-band magnitude in \cite{Liang:2014},
which is also derived using the NASA-Sloan Atlas sample.

\begin{figure}
	\centering
	\includegraphics[scale=0.45]{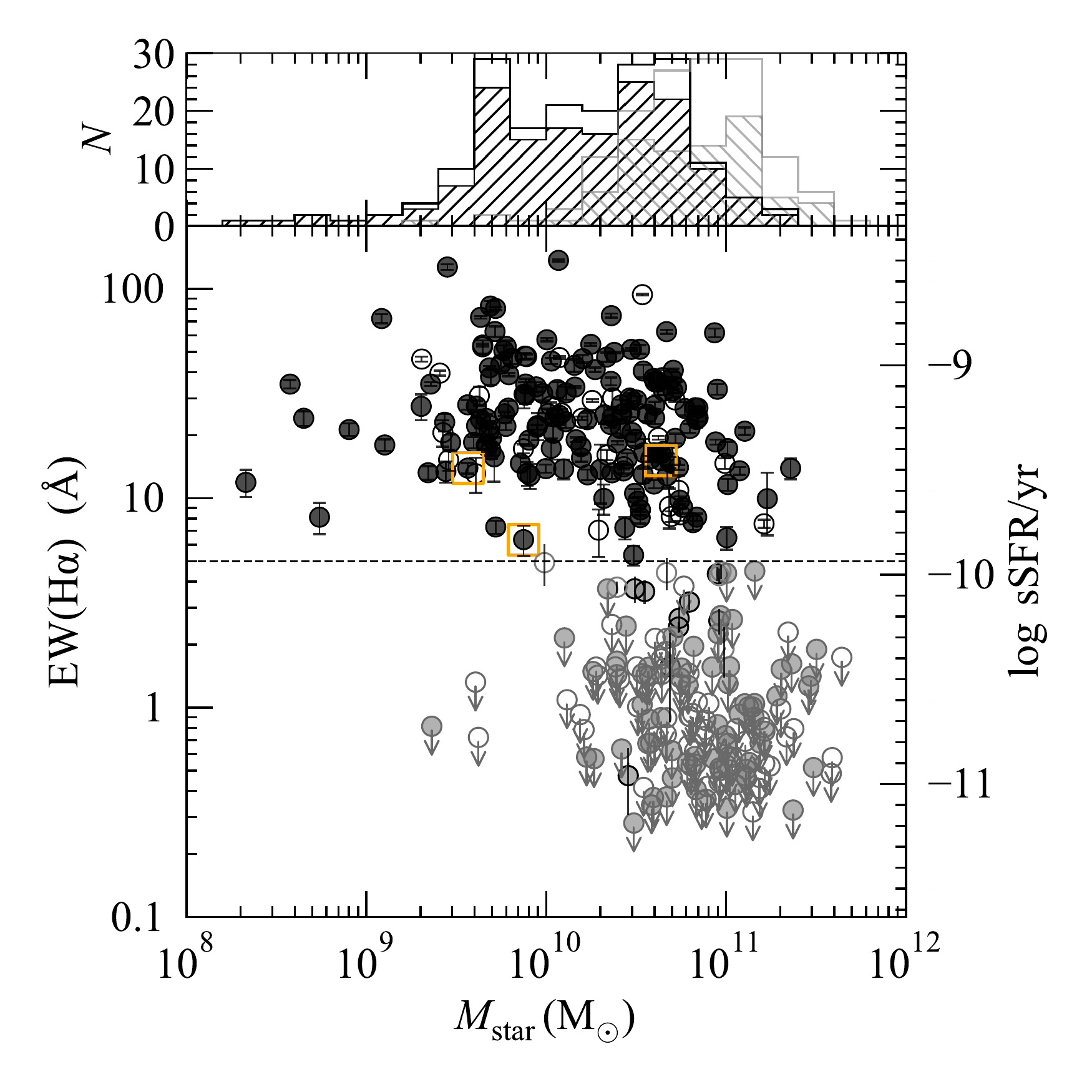}
	\caption{
	Correlation between 
	rest-frame \Halpha\ equivalent width and stellar mass.
	Filled and open circles represent respectively the 
	isolated and non-isolated galaxies.
	Galaxies likely to be dominated by an AGN 
	are marked in orange squares.
	Non-detections are shown as 2$\sigma$ upper 
	limits marked by light grey points
	with downward arrows.
	The vertical errors represent 1$\sigma$ uncertainties.
	The division between star-forming and quiescent galaxies 
	is shown as horizontal dashed line (EW(\Halpha)=5\AA). 
	Galaxies with $2\sigma$ upper limits greater than EW(\Halpha)=5\AA\ 
	do not have the sensitivity to distinguish 
	between star-forming and quiescent
	galaxies.  They are not displayed and 
	are excluded in our final sample.
	} 
	\label{figure:Mstar_SFR}
\end{figure}

We measure the equivalent width of the \Halpha\ emission 
line for each galaxy spectrum,
adopting the window definitions from \cite{Yan:2006}.
The stellar mass as a function of \Halpha\ equivalent width, 
EW(\Halpha), is presented in Figure \ref{figure:Mstar_SFR}.  
We also show on the right axis the inferred specific star-formation 
rate (sSFR)  following Equations (2) \& (4) in \cite{Fumagalli:2012}.
We mark cyan squares around galaxies likely to be dominated 
by an AGN (active galactic nucleus), based on the classification scheme
derived by \cite{Kewley:2001} with the optical 
line ratios \NII/\Halpha\ and \OIII/\Hbeta.
We find that there seems to be a clear distinction at 
EW(\Halpha) $\approx$ 5 \AA. Below EW(\Halpha)=5 \AA, 
only 15 out of 146 galaxies (10\%) have detected \Halpha\ emission
 at a 2-$\sigma$ level.  At EW(\Halpha)$>$5 \AA, most galaxies 
 (98\%) have detected \Halpha, except for 3 galaxies with low S/N spectra.
We therefore divide our galaxies into ``star-forming" 
and ``quiescent" galaxy samples using the criterion of 
whether their EW(\Halpha) are greater than 5 \AA.
A total of 327 out of 380 galaxies have high enough S/N 
measurements on the \Halpha\ emission, where 184 (143) 
are classified as star-forming (quiescent) galaxies.
The cut essentially limits our galaxy sample to $z\lesssim0.5$, 
beyond which the observed \Halpha\ emission is red-shifted to 
$\gtrsim 1 \rm \mu m$ and falls outside of the wavelength 
range of the optical spectrographs.
We find that our full sample spans a wide range in stellar mass,
from $M_{\rm star}=2\times10^8\,M_\odot$ to 
$M_{\rm star}=4\times10^{11}\,M_\odot$
with a median $\left<M_{\rm star}\right>=4\times10^{10}\,M_\odot$.
The star-forming galaxies have lower stellar mass from 
$\log M_{\rm star}/\msun=8.3-11.4$
with a median of $\log M_{\rm star}/\msun=10.3$, 
while the quiescent galaxy sample has higher
stellar mass from $\log M_{\rm star}/\msun=9.3-11.6$ 
with a median of $\log M_{\rm star}/\msun=10.8$.

\begin{figure*}
	\centering
	\includegraphics[scale=0.45]{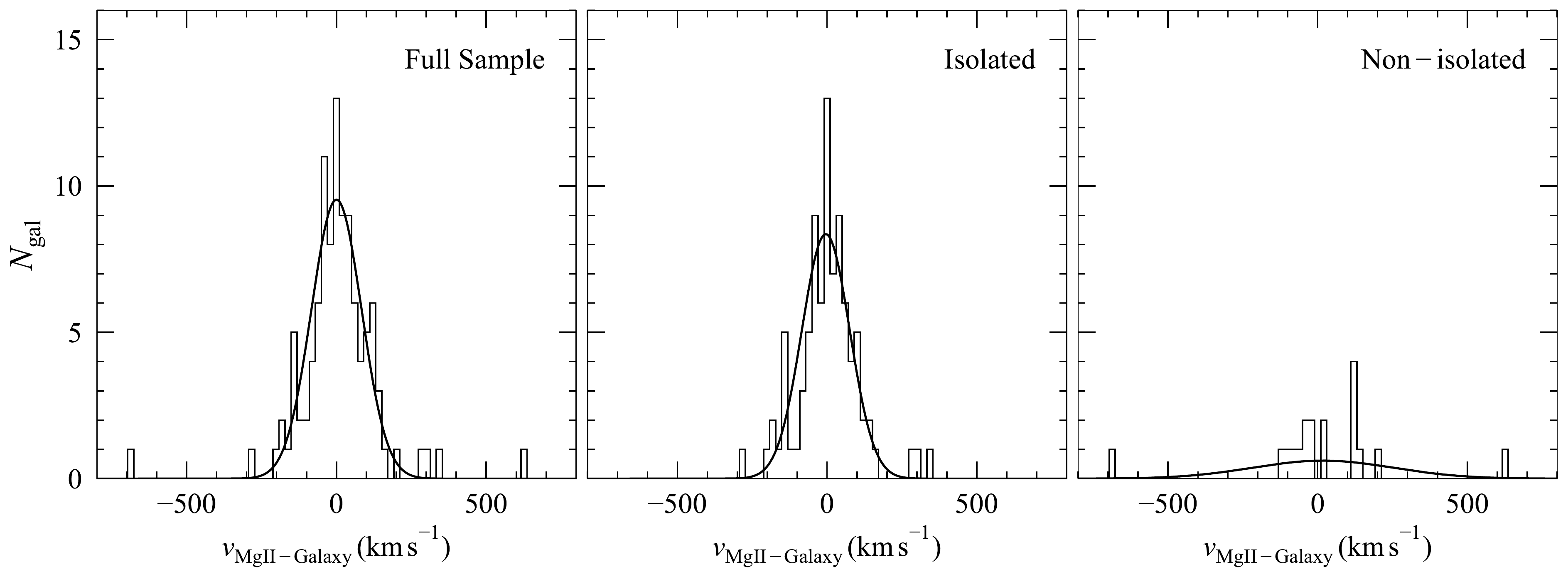}
	\caption{
	Relative velocity distributions of \MgII absorbers with respect
	to the galaxy systematic redshifts.
	The full galaxy sample is shown in the {\it left} panel and
	are separated into isolated and non-isolated galaxy samples
	in the {\it middle} and {\it right} panels.
	We detect associated \MgII absorbers in 103 out of 211 isolated galaxies
	and 18 out of 43 group systems searched.
	We characterize each relative velocity distribution by
	a single Gaussian profile (solid curve in each panel), excluding
	outliers according to an iterative 3$\sigma$ clipping.
	The full galaxy sample is well characterized by a Gaussian
	distribution with a mean velocity difference 
	$\langle v_{{\rm MgII-Galaxy}}\rangle=0$ \kms\ and
	dispersion of $\sigma$ = 84 \kms ({\it left} panel).	
	While adopting a Gaussian profile to characterize the velocity
	distribution of \MgII gas around isolated (non-isolated) galaxies 
	leads to a Gaussian profile centered at 
	$\langle v_{{\rm MgII-Galaxy}}\rangle=-4(17)$ \kms\ and
	$\sigma$ = 80 (235) \kms.  Since only 18 systems are 
	in the non-isolated sample,
	we also calculate the standard deviation of the velocity 
	dispersion to be $\sigma_{v\rm ,std} = 235$\kms, 
	consistent with what we obtain using the Gaussian 
	profile fitting. Note that for non-isolated systems, 
	the galaxy systematic redshift is defined
	as the luminosity-weighted redshift of member galaxies.
	}
	\label{figure:vdiff}
\end{figure*}

\begin{figure*}
	\centering
	\includegraphics[scale=0.45]{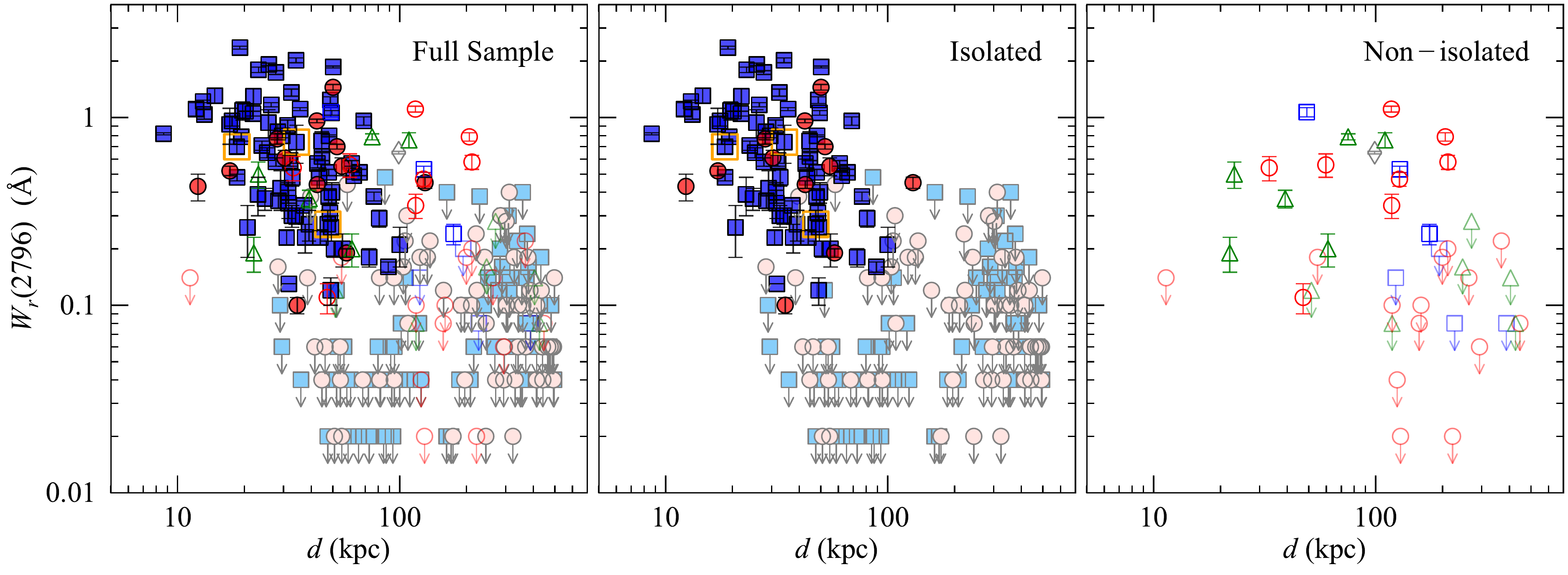}
	\caption{Rest-frame absorption equivalent width \ewr\ versus
	projected distance $d$ for the full ({\it left}), 
	isolated ({\it middle}; filled symbols) and non-isolated 
	({\it right}; open symbols) galaxy samples.
	Star-forming and quiescent galaxies are shown in blue 
	squares and red circles respectively. 
	Galaxies likely to be dominated by an AGN are marked in orange squares.
	Galaxies with detected Mg II absorbers 
	are shown in solid symbols with errorbars representing 
	measurement uncertainties. 
	Non-detections are displayed as 2$\sigma$ upper limits 
	(points with downward arrows) and lighter coloring.  
	Note that for non-isolated systems, we use the luminosity-weighted 
	projected distance as $d$ and display a single data point for each system.  
	Non-isolated systems which contain both star-forming and quiescent 
	galaxies are shown in green triangles.  We display in grey diamonds 
	for non-isolated systems which we cannot determine whether 
	their member galaxies are star-forming or quiescent.} 
	\label{figure:EW1}
\end{figure*}

\subsection{Absorber Properties}
For each galaxy in our sample, we searched for the
corresponding \MgII absorption doublet in the echellette
spectra of the background QSO within a radial velocity difference
$\Delta v=\pm1000$ \kms of the galaxy redshift.
When a \MgII absorber was identified, we measured its
rest-frame equivalent width and associated error by direct integration over
the line region in our continuum normalized spectrum.
We accepted the absorption lines according to a 2$\sigma$ detection threshold,
which is appropriate since the searches are performed at known
galaxy redshifts.
We then determined the
absorber redshift based on the best-fit line centroid of 
a Gaussian profile analysis of \MgII$\lambda2796$.
In cases where no \MgII features are detected, we placed
2$\sigma$ upper limits on the \MgII $\lambda2796$ 
equivalent widths. For \enquote{non-isolated} systems, 
we calculate the luminosity-weighted projected
position and redshift using members in each system 
and obtain its associated absorber properties following the same procedure.

In summary, the procedure yielded 85 physical galaxy-\MgII
pairs and 126 upper limits in the vicinities of 211 isolated galaxies, 
and 18 \MgII absorbers and 25 upper limits around 43 
\enquote{non-isolated} systems.  We were not able to obtain 
significant constraints for \MgII absorption equivalent widths 
around 14 galaxies and 2 \enquote{non-isolated} systems, where 
the expected QSO spectra are contaminated by other strong
absorption features (e.g. \CIV $\lambda\lambda$ 1548, 1550)  
or the atmosphere O3 absorption complex at $\lambda \approx 3200$.
Also, six galaxies and two \enquote{non-isolated} systems were 
found at spectroscopic redshifts $z_{\rm spec}\lesssim0.09$,
falling outside the wavelength range of the MagE spectrograph. 
We present the properties of each spectroscopically confirmed
galaxy sample in Table \ref{table:absorberSummary} (Columns $1-10$),
and the associated redshift and absorption equivalent width of \MgII absorbers
in Columns 11 and 12 of Table \ref{table:absorberSummary}.

\begin{table*}
\caption{Galaxies and Absorption Systems} 
\label{table:absorberSummary} 
\resizebox{\textwidth}{!}{\begin{tabular}{lrrcccccccccc}
\hline \hline
\multicolumn{10}{c}{Galaxies} & & \multicolumn{2}{c}{Absorption Systems} \\
\cline{1-10}
\cline{12-13}
& \multicolumn{1}{c}{$\Delta \alpha$} & \multicolumn{1}{c}{$\Delta \delta$} & &
\multicolumn{1}{c}{$d$} & & & & $R_{\rm h}$ & EW(\Halpha) & & & {$W(2796)$} \\
\multicolumn{1}{c}{ID} & \multicolumn{1}{c}{(arcsec)} &
\multicolumn{1}{c}{(arcsec)} & {$z_{\rm gal}$} & {(kpc)} & {$r'$} &
{$M_B$} & $\log\,\mstar/\msun$ & (kpc) & (\AA) & & {$z_{\rm abs}$} & {(\AA)} \\
\multicolumn{1}{c}{(1)} & \multicolumn{1}{c}{(2)} & \multicolumn{1}{c}{(3)} &
{(4)} & {(5)} & {(6)} & {(7)} & {(8)} &  {(9)} & {(10)} & & {(11)} & {(12)}  \\
\hline
SDSSJ001336.14+141428.04 & $5.8$ & $3.9$ & 0.1910 & 21.9 & 19.52 & $-19.93$ & 9.7 & 143.9 & 20.8 $\pm$ 1.2 & & 0.1908 & 1.30 $\pm$ 0.13 \nl
SDSSJ003009.52+011445.25 & $-65.9$ & $40.1$ & 0.1845 & 238.7 & 17.72 & $-21.51$ & 10.7 & 247.7 & 40.9 $\pm$ 0.7 & & 0.1845 & $<$ 0.10 \nl
SDSSJ003010.18+011219.92 & $-55.9$ & $-105.2$ & 0.1501 & 311.6 & 18.48 & $-20.14$ & 10.4 & 200.6 & 49.7 $\pm$ 0.4 & & 0.1501 & $<$ 0.13 \nl
SDSSJ003014.16+011359.19 & $3.9$ & $-6.0$ & 0.2125 & 24.7 & 20.81 & $-18.59$ & 9.7 & 142.9 & 19.4 $\pm$ 4.0 & & 0.2126 & 0.37 $\pm$ 0.05 \nl
SDSSJ003339.86$-$005522.39 & $-5.4$ & $3.3$ & 0.2124 & 21.9 & 18.99 & $-20.79$ & 9.8 & 149.0 & 46.7 $\pm$ 2.9 & & 0.2121 & 1.05 $\pm$ 0.03 \nl
SDSSJ003406.33$-$085448.74 & $-15.4$ & $3.4$ & 0.1403 & 38.5 & 18.04 & $-20.18$ & 10.6 & 228.1 & $<$ 0.7 & & 0.1403 & $<$ 0.14 \nl
SDSSJ003407.78$-$085453.28 & $6.5$ & $-1.1$ & 0.3617 & 32.8 & 21.59 & $-19.43$ & 9.8 & 134.7 & 50.7 $\pm$ 3.0 & & 0.3616 & 0.48 $\pm$ 0.05 \nl
SDSSJ003411.94$-$005808.52 & $-16.5$ & $138.3$ & 0.1539 & 372.2 & 19.20 & $-19.77$ & 9.7 & 149.0 & 80.7 $\pm$ 1.4 & & 0.1539 & $<$ 0.14 \nl
SDSSJ003412.85$-$010019.81 & $-2.9$ & $7.1$ & 0.2564 & 30.4 & 20.08 & $-19.79$ & 10.3 & 184.1 & $<$ 3.7 & & 0.2564 & 0.61 $\pm$ 0.06 \nl
SDSSJ003414.49$-$005927.51 & $21.8$ & $59.3$ & 0.1202 & 136.9 & 17.27 & $-20.54$ & 10.7 & 267.4 & $<$ 1.4 & & 0.1212 & $<$ 0.21 \nl
\hline
\multicolumn{11}{c}{non-isolated Galaxies}\\
\hline
SDSSJ000548.29$-$084757.25 & $0.7$ & $11.2$ & 0.3293 & 53.2 & 19.21 & $-21.45$ & 11.0 & 297.9 & 1.9 $\pm$ 0.5 & & 0.3288 & 0.11 $\pm$ 0.02 \nl
SDSSJ000548.59$-$084801.16 & $5.2$ & $7.3$ & 0.3293 & 42.4 & 19.43 & $-21.22$ & 10.8 & 251.4 & $<$ 0.6 & & ... & ... \nl
SDSSJ003339.66$-$005518.39 & $-8.2$ & $7.2$ & 0.1760 & 32.5 & 19.66 & $-18.96$ & 10.1 & 172.4 & $<$ 1.1 & & 0.1759 & 0.19 $\pm$ 0.04 \nl
SDSSJ003341.48$-$005522.82 & $19.1$ & $2.8$ & 0.1758 & 57.4 & 20.79 & $-18.32$ & 9.4 & 133.1 & 20.6 $\pm$ 3.0 & & ... & ... \nl
SDSSJ003357.90$-$085356.03 & $-141.7$ & $56.1$ & 0.1381 & 367.9 & 17.76 & $-20.50$ & 10.6 & 235.0 & $<$ 0.7 & & 0.1381 & $<$ 0.22 \nl
SDSSJ010133.65$-$005028.98 & $-32.9$ & $-19.9$ & 0.2605 & 154.9 & 17.98 & $-21.97$ & 11.2 & 409.4 & $<$ 0.8 & & 0.2605 & $<$ 0.10 \nl
SDSSJ012557.41$-$000749.58 & $-83.7$ & $32.9$ & 0.2574 & 359.3 & 18.66 & $-21.20$ & 11.0 & 302.5 & $<$ 0.4 & & 0.2569 & 0.79 $\pm$ 0.04 \nl
SDSSJ012603.20$-$000827.82 & $3.2$ & $-5.3$ & 0.2571 & 24.6 & 18.55 & $-21.44$ & 10.8 & 255.4 & $<$ 1.5 & & ... & ... \nl
SDSSJ020111.79$-$002537.67 & $-26.1$ & $106.3$ & 0.1568 & 296.8 & 19.44 & $-19.46$ & 9.9 & 158.1 & 47.7 $\pm$ 1.6 & & ... & ... \nl
SDSSJ020117.91$-$002559.59 & $65.8$ & $84.4$ & 0.1577 & 291.5 & 18.97 & $-19.63$ & 10.3 & 189.6 & $<$ 1.4 & & ... & ... \nl
SDSSJ020119.90$-$002831.16 & $95.5$ & $-67.2$ & 0.1602 & 322.4 & 17.83 & $-20.75$ & 10.8 & 271.5 & $<$ 0.7 & & 0.1602 & $<$ 0.15 \nl
\hline \hline
	\multicolumn{7}{l}{The full table is available in the on-line version of the paper.}
	\end{tabular}}
\end{table*}

The final isolated galaxy-absorber pair sample spans a projected distance
range of $d=9-497$ kpc with a median of $\langle d \rangle_{\rm med}=73$ kpc.
The redshifts of the isolated galaxies range from $z=0.10-0.48$ with
a median of $\langle z \rangle_{\rm med}=0.21$, and the rest-frame
absolute $B$-band magnitudes range from $M_B= -16.7$ to  
$M_B=-22.8$ with a median of $\langle M_B \rangle_{\rm med}=-20.5$.
The \enquote{non-isolated} systems span a redshift range of 
$z=0.12-0.47$ with a median of $\langle z \rangle_{\rm med}=0.19$,
with projected distance from $d=11-446$ kpc with a median of 
$\langle d \rangle_{\rm med}=128$ kpc.

We examine the relative velocity distribution of \MgII absorbers
with respect to the systematic redshifts of the galaxies.  In the left panel
of Figure \ref{figure:vdiff} we present the velocity dispersion of the
detected \MgII absorbing gas around the full galaxy-absorber pair sample.
We also show the velocity dispersion separately for isolated and
non-isolated galaxy-absorber pairs in the central and right panels of
Figure \ref{figure:vdiff}.
We characterize the velocity distribution using a Gaussian profile
with iterative 3-sigma clipping to exclude outliers.
The velocity distribution of \MgII absorbing gas around galaxies 
can be characterized by a single Gaussian distribution of mean 
velocity difference $\langle v_{\rm MgII-Galaxy}\rangle=0$\kms and dispersion 
$\sigma_v = 84$\kms (left-hand panel of Figure \ref{figure:vdiff}),
while the velocity distribution for
isolated galaxies is best represented by a Gaussian profile
centered at $\langle v_{\rm MgII-Galaxy}\rangle=-4$\kms and 
$\sigma_v = 80$\kms.
For \enquote{non-isolated} systems, we find the associated \MgII absorbers
have a broad velocity distribution with a standard deviation of 
$\sigma_{v\rm ,std} = 235$\kms.

We present in Figure \ref{figure:EW1} the correlation between the 
strength of \MgII absorption and galaxy projected distance.  
Following the presentation in Figure \ref{figure:vdiff},
we show the distributions separately for isolated and non-isolated galaxies
to investigate the influence of galaxy environment.
Similar to previous surveys \citep[e.g.][]{Chen:2010}, 
we find a clear trend of decreasing absorption strength
with increasing projected distance for isolated galaxies 
(see the middle panel of Figure \ref{figure:EW1}).  
Beyond 70 kpc, no \MgII absorbing gas with 
\ewr $> 0.5\rm \AA$ {is} found,
and no detections are present beyond 
projected distance $d\gtrsim150$ kpc ($d\geq\hat R_{\rm gas}$).
In contrast, while non-isolated systems do have lower 
\MgII covering fraction at larger projected distance $d$, 
we do not find a clear anti-correlation between the \ewr\ and $d$. 
The strong \MgII absorbers of \ewr $\sim 0.5\rm \AA$ are 
detected out to 150 kpc.
{Note that using the projected distance of the nearest 
galaxy in non-isolated systems
as $d$ does not change the lack of anti-correlation 
in the \ewr\ versus $d$ plot.}

To determine whether the recent star formation of galaxies
has an impact on \MgII absorbing gas, we further divide the isolated
galaxy sample into star-forming and quiescent galaxy samples
and show the distributions of \ewr\ versus projected distance 
in the upper panels of Figure \ref{figure:fit_d}.
While both galaxy samples appear to occupy a similar \ewr\ versus
 $d$ space, qualitatively the star-forming galaxies show strong 
inverse correlation whereas only a modest trend is revealed for 
quiescent galaxies.

\section{Analysis}
\label{section:analysis}
In the previous section, we use the samples of 
211 isolated galaxies (73 quiescent galaxies and
138 star-forming galaxies) and 43 non-isolated galaxies
to show that the strength and incidence of \MgII absorbing
gas appear to depend on galaxy properties.
In this section, we quantify the correlation between \ewr\ and
different galaxy properties.  We obtain and assess various models that
well describe the data and present the best-fit results. 

\subsection{Fitting Procedure and Model Evaluation}
We perform a likelihood analysis to obtain the best-fit models and
to better characterize the correlation between galaxy properties and \MgII 
absorbers.  The generalized functional form to describe
the mean \MgII absorption equivalent width $\bar{W}_r(2796)$ is
    \begin{align}
       \bar{W}_r(2796) = f(x_1,x_2,...)
   \end{align}
where $x_i$'s are independent measurements of galaxy properties
including projected distance ($d$), rest-frame absolute $B-$band luminosity
 ($M_{\rm B}$) and stellar mass ($M_{\rm star}$).
We adopt a simple power-law profile to describe
the correlation between $W_r(2796)$, $d$ and other properties.
In logarithmic space, the model is expressed as a linear equation
\begin{equation}
\log\,\bar{W}_r(2796)\,=\,a_0 + a_1 \log\,d + a_2 X + ....
\label{eq:powerlaw}
\end{equation}
In addition, we introduce a non-parametric covering fraction 
$\epsilon$ in the model to describe
the clumpy nature of \MgII gaseous halos.
This is motivated by our findings that the covering fraction 
of \MgII absorbing gas may be less than unity and varying
at different projected distances.
As shown in Figure \ref{figure:EW1},   
a non-negligible fraction of galaxies at small projected 
distances ($d\lesssim 40$ kpc) do not give rise to \MgII 
absorption to sensitive upper limits, and we find an 
increasing fraction of upper limits at larger projected distances.
We divide the mean covering fraction $\epsilon_k \,(k=1-4)$ 
into four projected distance intervals.  To obtain a better sampling 
of $\epsilon$ at small projected distances, the first two bins ($k=1,2$)
are designed for $\rm d \lesssim 100\,kpc$, where most 
\MgII absorbers are found.  The other two bins ($k=3,4$) are used 
for larger projected distances, each with a roughly equal number of galaxies.

Here we perform the maximum likelihood analysis
to determine the values of coefficients $a_i$ and 
four mean covering fractions that best represent the data.
The likelihood function is defined as
  \begin{align}
 {\mathscr L} &= \prod_{i=1}^n \big\{ (\epsilon|d_i) g_{i}  + [1- (\epsilon|d_i)]h_{i} \big\}
 \label{eq:likelihood}
  \end{align}
where $g_i$ represents the probability density function that the
QSO sightline intercepts the \MgII gas around galaxy and its strength (\ewr)
follows the underlying power-law model, and $h_i$
denotes the pdf that the QSO sightline does not intercept 
any \MgII gas around galaxy $i$.
Here $\langle\epsilon|d_i\rangle$ is the non-parametric 
mean covering fraction given the projected distance of galaxy $i$ ($d_i$).
The two probability density functions $g_i$ and $h_i$ are 
weighted according to the mean
covering fraction ($\epsilon|d_i$) and combined to get 
the likelihood function of a single galaxy $i$. 

The first probability density function $g_i$ is expressed as
\begin{align}
   g_{i}  = \int_{0}^{\infty}  dW^\prime & \left(\frac{1}{\sqrt{2\pi}\ W^\prime\ \sigma_{\rm insc}} {\rm exp}\left\{-\frac{1}{2} \left[ \frac{\ln(W^\prime)-\ln(\bar{W})}{\sigma_{\rm insc}}\right]^2\right\}\right)  \nonumber   \\
      & \times \left( \frac{1}{\sqrt{2 \pi} \sigma_i} {\rm exp}\left\{-\frac{1}{2}\left(\frac{W^\prime - W_i}{\sigma_i}\right)^2\right\} \right)
      \label{eq:gi}
\end{align}
where $W_i$ is the observed $W_r(2796)$ for galaxy $i$, 
$\bar{W}$ is the model expectation, and $\sigma_i$ is 
the measurement error of $W_i$.
Specifically, the first term takes into account the 
intrinsic scatter ($\sigma_{\rm insc}$) of 
a given model expectation ($\bar{W}$) due to variations 
between individual galaxies and between different sightlines 
probing the same galaxy \citep[e.g.][]{Chen:2014}.
Motivated by Figure \ref{figure:EW1} and previous studies 
\citep[e.g.,][]{Chen:2010}, we model the intrinsic scatter as a 
constant in logarithmic space, independent of 
galaxy properties and projected distance.
The second term represents the pdf of a normal 
distribution induced by measurement uncertainty $\sigma_i$.

On the other hand, $h_i$ is defined simply as a normal 
distribution with measurement error $\sigma_i$ and a 
mean consistent with absence of gas (i.e. zero)
\begin{align}
    h_{i} = \frac{1}{\sigma_i \sqrt{2\pi}} {\rm exp}\left( -\frac{1}{2} \left[\frac{W_i - 0.0}{\sigma_i}\right]^2\right)
\end{align}
We note that for each non-detection, we also measure the rest-frame
equivalent width by direct summation of the continuum 
normalized spectrum over a resolution element,
centering at the systematic redshift of the galaxy.
This allows us to appropriately exploit the constraints
 from both detections and non-detections the same way.
We multiply the likelihood function of each galaxy $i$ over 
a total of $n$ target galaxies to obtain the total likelihood 
function in Equation \ref{eq:likelihood}. We assess the 
confidence intervals of derived model parameters using 
the Markov Chain Monte Carlo (MCMC) method.

\begin{figure*}
	\centering
	\includegraphics[clip,scale=0.75]{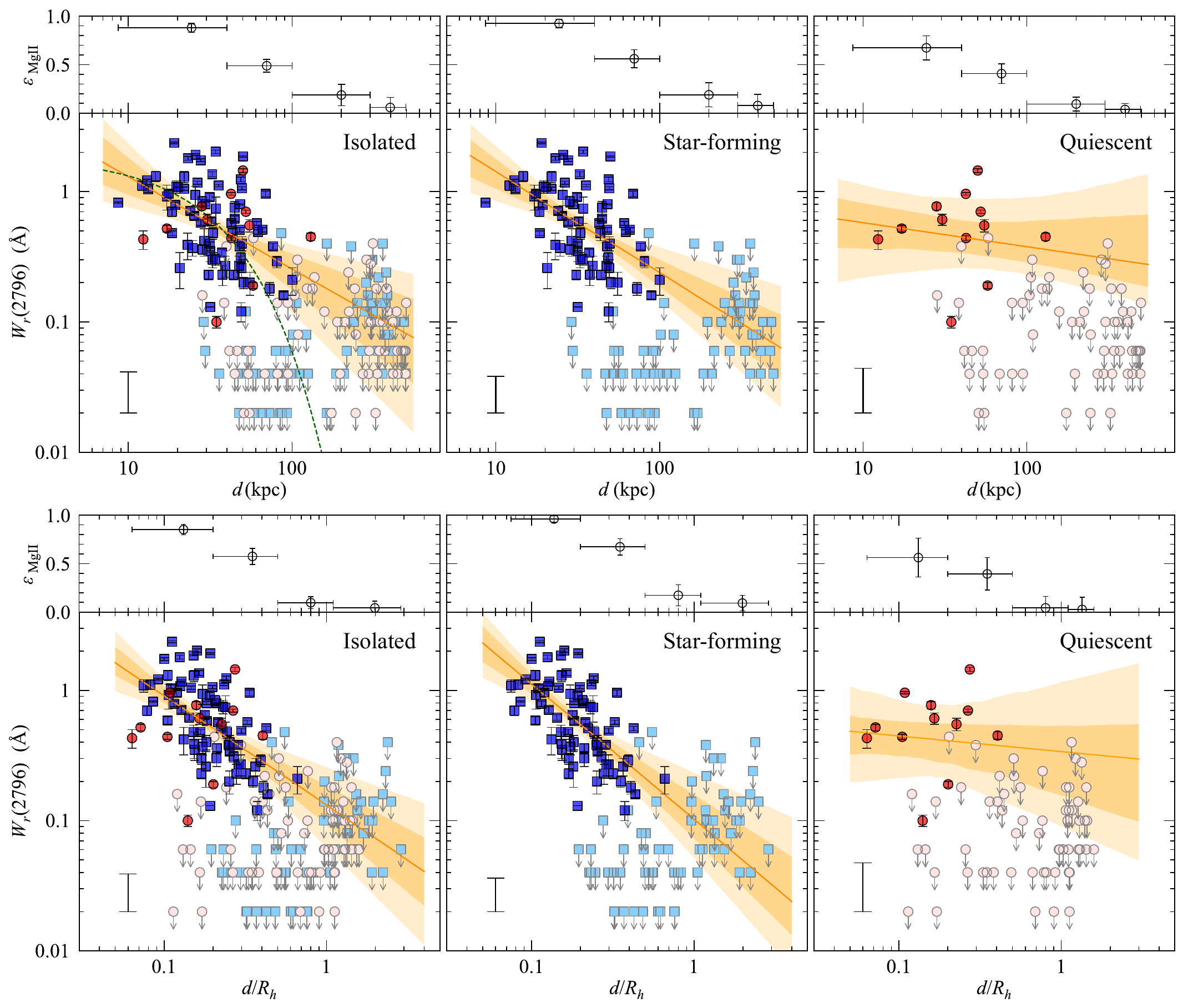}
	\caption{Rest-frame equivalent width \ewr\ versus 
	projected distance $d$ ({\it upper panels}) and $R_{\rm h}$-normalized 
	projected distance $d/R_{\rm h}$ ({\it lower panels}).
	The full isolated galaxy sample is shown in the {\it left} panels,  
	and are separated into isolated star-forming 
	and quiescent galaxy samples in the {\it middle} and {\it right} panels.	
	Symbols are the same as in Figure \ref{figure:EW1}.
	The orange solid lines represent the best-fit power-law models.
	The dark and light shaded bands represent the 68\% and 95\%
	confidence intervals from the Markov chain Monte Carlo realizations. 	
	The best-fit intrinsic scatter is marked in the lower left corner.
	At the {\it top} of each panel, we show the best-fit results of 
	the 4 non-paramatric mean covering fractions
	from the maximum likelihood analysis.  The horizontal bars mark the
	full range of projected distance within each bin and vertical error bars
	represent the 68\% confidence interval.
	The best-fit models are discussed in Equation 
	(\ref{equation:bestfitfull})$-$ Equation (\ref{equation:bestfitRhred}) 
	and in the text. For comparison, the green dashed curve in the 
	{\it upper left} panel shows the log-linear maximum 
	likelihood fit from \citet{Nielsen:2013a}.}
	\label{figure:fit_d}
\end{figure*}

\subsection{Dependence of Extended Gas on Galaxy Projected Distance}
First, we seek the best-fit models to describe the
dependence of \ewr\ on $d$ (see the upper left panel 
of Figure \ref{figure:fit_d}; Model I).
The maximum likelihood solution for isolated galaxies is 
\begin{align}
\log\,\bar{W}_r(2796)\, &=\, (0.83\pm0.38) -(0.72\pm0.25) \log\,d 
\label{equation:bestfitfull}
\end{align}
with an intrinsic scatter $\sigma=0.32\pm0.04$ in common logarithm.
Note that $\sigma$ is a simple conversion of 
$\sigma_{\rm insc}$ (see Equation \ref{eq:gi})
from natural to common logarithm.
The errors in the coefficients are 1$\sigma$ uncertainties.
The non-parametric covering fractions are 
$\epsilon_1=0.87\pm 0.05$,
$\epsilon_2=0.49\pm 0.06$,
$\epsilon_3=0.19\pm 0.12$, and
$\epsilon_4=0.06\pm 0.06$.
The intervals of the covering fraction bins are listed in 
Table \ref{table:LikelihoodSummary}.
The maximum likelihood solution shows a significant 
anti-correlation ($\sim\,3\sigma$ level)
between the \MgII absorption strength and projected distance $d$.
Our results also show that the mean gas covering fraction
declines steeply as the projected distance increases.
{We note that the inferred low covering fraction at 
$d\gtrsim100\,$kpc may be driven
by weak absorbers (i.e. \ewr$\lesssim$0.04\AA) for 
which our data have insufficient S/N to uncover. }

To understand the influence of galaxy types on their \MgII 
absorbing gas properties, we also obtain the best-fit models
for isolated star-forming and quiescent galaxy samples 
(respectively the upper middle and right panels of Figure \ref{figure:fit_d}).
For star-forming galaxies, we find based on the likelihood
analysis a best-fit model
\begin{align}
\log\,\bar{W}_r(2796)\, &=\, (0.94\pm0.30) -(0.78\pm0.20) \log\,d 
\label{equation:bestfitblue}
\end{align}
with an intrinsic scatter $\sigma=0.28\pm0.03$ 
and mean covering fractions of 
$\epsilon_1=0.92\pm 0.05$,
$\epsilon_2=0.56\pm 0.09$,
$\epsilon_3=0.19\pm 0.12$, and
$\epsilon_4=0.08\pm 0.08$.
Similar to the full sample, the \ewr\ and incidence of gas 
for star-forming galaxies both decrease as increasing projected distance.
For quiescent galaxies, however, the dependence of
\MgII absorbing gas and projected distance reveals a stark contrast.
The best-fit power-law model from the likelihood analysis yields
\begin{align}
\log\,\bar{W}_r(2796)\, &=\, (-0.17\pm0.20) -(0.08\pm0.34) \log\,d 
\label{equation:bestfitred}
\end{align}
with $\sigma=0.34\pm0.06$ and covering fractions of 
$\epsilon_1=0.67\pm 0.11$,
$\epsilon_2=0.41\pm 0.10$,
$\epsilon_3=0.09\pm 0.07$, and
$\epsilon_4=0.04\pm 0.04$.
While the best-fit mean covering fractions decline with increasing $d$,
they are $\sim30\%$ lower compared to that of the star-forming 
ones at $d<300$ kpc. Furthermore, the maximum likelihood 
solution shows no statistically significant dependence between 
\MgII absorption strength and $d$ among detections.

In Figure \ref{figure:Mstar_SFR}, we show that the star-forming 
galaxy sample spans a wide range of stellar masses from 
$M_{\rm star}=10^8-10^{11.3}$, while quiescent galaxies
have on average higher stellar masses ranging from 
$M_{\rm star}=10^{9.3}-10^{11.6}$.
It has been shown in observations \citep[e.g.,][]{Chen:2010b} 
and theoretical models \citep[e.g.,][]{Mo:1996,Maller:2004}
that more massive galaxies tend to have more extended gaseous halos.
To address whether the apparent differential observed 
$\bar{W}_r(2796)$ versus $d$ plane between star-forming and 
quiescent galaxy samples can be affected by the gaseous 
halos with various sizes, we estimate the dark matter 
halo radius ($R_{\rm h}$) of individual galaxies and examine 
how \MgII absorption strength varies with
$R_{\rm h}$-normalized projected distance ($d/R_{\rm h}$).
The halo radius $R_{\rm h}$ is calculated following the 
prescription in \cite{Liang:2014}.  Specifically, we obtain 
halo mass ($M_{\rm h}$) using the stellar-mass-to-halo-mass 
relation derived in \cite{Kravtsov:2018}, and $M_{\rm h}$ is then 
converted to $R_{\rm h}$ with standard cosmology \citep{Bryan:1998}.

We present the correlation between $\bar{W}_r(2796)$ 
and galaxy $R_{\rm h}$-normalized projected distance ($d/R_{\rm h}$)
in the lower panels of Figure \ref{figure:fit_d} (Model II).
We note that \MgII absorbing gas declines steeply beyond 
0.4 $R_{\rm h}$. Essentially only one galaxy at $d=0.66\,R_{\rm h}$ 
has an associated \MgII absorber.
The maximum likelihood solution for the full isolated galaxy sample is

\begin{align}
\log\,\bar{W}_r(2796)\, &=\, (-0.88\pm0.15) -(0.84\pm0.20) \log[d/R_{\rm h}]
\label{equation:bestfitRhFull}
\end{align}
with an intrinsic scatter $\sigma _c=0.29\pm 0.03$ and 
covering fractions of 
$\epsilon_1=0.85\pm 0.06$,
$\epsilon_2=0.57\pm 0.08$,
$\epsilon_3=0.10\pm 0.06$, and
$\epsilon_4=0.07\pm 0.07$.

We find that after accounting for the mass scaling of 
gaseous radius, the slope of the anti-correlation is
steepened by 17\% and becomes more statistically significant.  
For the isolated star-forming galaxies, we obtain a similar 
improvement based on the likelihood analysis

\begin{align}
\log\,\bar{W}_r(2796)\, &=\, (-0.98\pm0.17) -(1.03\pm0.22) \log[d/R_{\rm h}]
\label{equation:bestfitRhblue}
\end{align}
with an intrinsic scatter $\sigma _c=0.26\pm 0.03$ and 
covering fractions of 
$\epsilon_1=0.95\pm 0.04$,
$\epsilon_2=0.67\pm 0.08$,
$\epsilon_3=0.18\pm 0.10$, and
$\epsilon_4=0.12\pm 0.12$.
The best-fit result also shows a steeper slope by 32\% 
and a $\sim 5\sigma$ significance of the anti-correlation.
In contrast, including scaling with $R_{\rm h}$ does not 
improve the observed $\bar{W}_r(2796)$ versus $R_{\rm h}$ 
anti-correlation for the quiescent galaxy sample.
We find a best-fit model of

\begin{align}
\log\,\bar{W}_r(2796)\, &=\, (-0.48\pm0.27) -(0.12\pm0.24) \log[d/R_{\rm h}]
\label{equation:bestfitRhred}
\end{align}
with an intrinsic scatter $\sigma _c=0.37\pm 0.09$ and 
covering fractions of 
$\epsilon_1=0.53\pm 0.18$,
$\epsilon_2=0.39\pm 0.15$,
$\epsilon_3=0.07\pm 0.07$, and
$\epsilon_4=0.05\pm 0.05$.
The observed \MgII absorption strength versus $d/R_{\rm h}$ 
remains consistent with a flat distribution.

On the other hand, previous studies have shown that galaxy 
$B$-band luminosity, which is a direct observable,
also has an impact on both the extent of \MgII absorbing gas
and absorption strength \citep[e.g.][]{Chen:2008,Chen:2010}.
Therefore, we include rest-frame $B$-band magnitude $M_B$ 
in the power-law model ($X\equiv M_B - M_{B_*}$ in 
Equation \ref{eq:powerlaw}) to examine how \MgII absorber 
strength scales with $M_B$ (Model III).
We adopt $M_{B_*}=-20.6$ from \cite{Faber:2007},
the characteristic $B$-band magnitude for describing blue 
galaxy population at $z\sim0.4$.
Based on the maximum likelihood analysis, we find the 
\MgII gaseous extent scales with $d$ and $M_B$ following 
\begin{align}
\log\,\bar{W}_r(2796)\, = &\, (1.22\pm0.25)-(0.94\pm0.16) \log\,d \nonumber \\
& - (0.09\pm0.03) \times (M_{\rm B} - M_{\rm B*})
\label{equation:bestfitMBAll}
\end{align}
with an intrinsic scatter of $\sigma _c=0.28\pm 0.03$ and 
covering fractions of 
$\epsilon_1=0.95\pm 0.04$,
$\epsilon_2=0.58\pm 0.09$,
$\epsilon_3=0.11\pm 0.07$, and
$\epsilon_4=0.04\pm 0.04$.
The results show that including the intrinsic $B$-band luminosity
also leads to a steeper slope of the anti-correlation between \ewr\ and $d$.
Note that $M_B$ has a negative correlation coefficient, consistent
with the expectation that brighter galaxies have larger extent of gas.

For star-forming galaxies, we also perform the likelihood analysis and find
the observed absorber strength is best described by
\begin{align}
\log\,\bar{W}_r(2796)\, = &\, (1.57\pm0.28)-(1.14\pm0.18) \log\,d \nonumber \\
& - (0.12\pm0.02) \times (M_{\rm B} - M_{\rm B*})
\label{equation:bestfitMBblue}
\end{align}
with an intrinsic scatter $\sigma _c=0.25\pm 0.03$ and 
covering fractions of 
$\epsilon_1=0.97\pm 0.02$,
$\epsilon_2=0.60\pm 0.08$,
$\epsilon_3=0.18\pm 0.11$, and
$\epsilon_4=0.05\pm 0.05$.
The results show trends similar to the full isolated galaxy sample.
We find a steeper anti-correlation between \ewr\ versus $d$
and an anti-correlation between absorption strength and $M_B$.
The intrinsic scatter is slightly smaller than the results obtained
without the scaling of $B$-band luminosity.

For quiescent galaxies, we obtain a best-fit model of
\begin{align}
\log\,\bar{W}_r(2796)\, = &\, (0.09\pm0.33)-(0.27\pm0.20) \log\,d \nonumber \\
& - (0.02\pm0.05) \times (M_{\rm B} - M_{\rm B*})
\label{equation:bestfitMBred}
\end{align}
with an intrinsic scatter $\sigma _c=0.34\pm 0.06$ and 
covering fractions of 
$\epsilon_1=0.54\pm 0.17$,
$\epsilon_2=0.37\pm 0.17$,
$\epsilon_3=0.07\pm 0.07$, and
$\epsilon_4=0.03\pm 0.03$.
Contrary to the star-forming galaxies, we do not find
strong correlation between \ewr\ and $d$.
A slight negative correlation between \ewr\ and $M_B$ is shown.
The addition of $M_B$ scaling in the power-law model does
not seem to reduce the intrinsic scatter of the \ewr\ versus $d$ relation.

\begin{figure*}
	\centering
	\includegraphics[clip,scale=0.75]{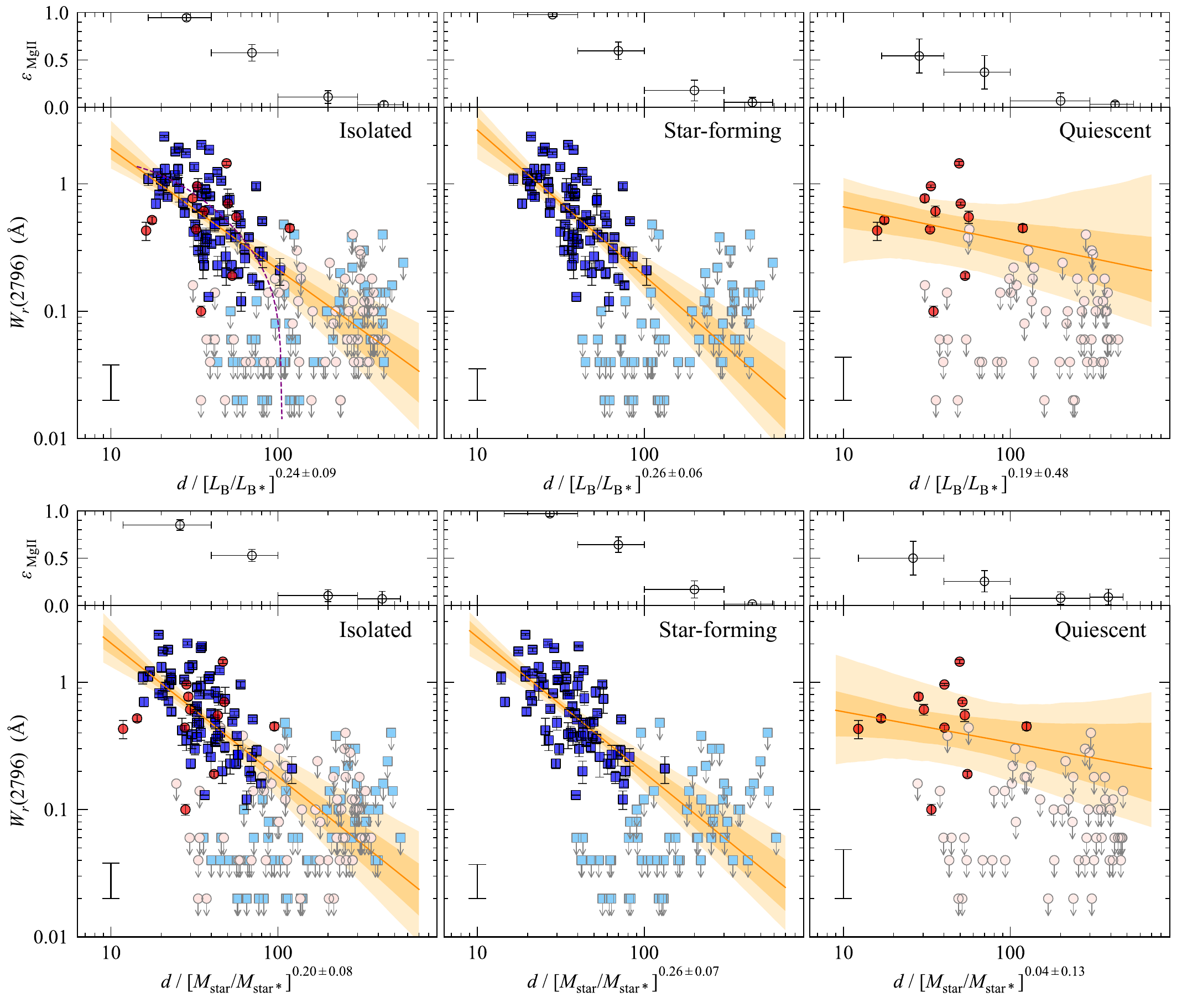}
	\caption{
	Similar to Figure \ref{figure:fit_d}, but the correlation between \ewr\ versus 
	$d$ also accounts for the scaling with galaxy
	$B$-band luminosity in the {\it upper} panels and 
	stellar mass in the {\it lower} panels.  
	Note that the scaling coefficients $a^{\prime}\equiv 2.5\times(a_2/a_1)$ in
	$d^\prime=d/[L_{\rm B}/L_{\rm B*}]^{a^{\prime}}$ 
	and $a^{\prime \prime}\equiv -(a_2/a_1)$ in
	$d^\prime=d/[M_{\rm star}/M_{\rm star*}]^{a^{\prime \prime}}$  
	are calculated based on the
	best-fit models in Equation (12)$-$(17).
	For comparison, the purple dashed curve in the {\it upper left} 
	panel represents the best-fit isothermal model from \citet{Chen:2010}. }
	\label{figure:fit_MB}
\end{figure*}

It is notable that although $M_B$ is known to scale
with halo mass \citep[e.g.,][]{Zheng:2007}, it also 
correlates with \OII\ luminosity \citep[e.g.,][]{Zhu:2009}
and thus might be coupled with recent star-formation. 
Here we inspect the correlation between \ewr\ and total stellar
mass $M_{\rm star}$, which is believed to be
a good tracer of halo mass \citep{More:2011} (Model IV). 

We obtain the best-fit models with the scaling of total stellar
mass $M_{\rm star}$ of
\begin{align}
\log\,\bar{W}_r(2796)\, = &\, (1.35\pm0.25)-(1.05\pm0.17) \log\,d \nonumber \\
& + (0.21\pm0.08) \times (M_{\rm star} - M_{\rm star*})
\label{equation:bestfitMstarAll}
\end{align}
where $\log M_{\rm star*}/\msun=10.3$.
The best-fit intrinsic scatter is $\sigma _c=0.28\pm 0.03$ and 
best-fit covering fractions are 
$\epsilon_1=0.85\pm 0.06$,
$\epsilon_2=0.53\pm 0.07$,
$\epsilon_3=0.10\pm 0.07$, and
$\epsilon_4=0.07\pm 0.07$.
The best-fit coefficients show a positive correlation between
\ewr\ and $M_{\rm star}$, and an anti-correlation between
\ewr\ and $d$, consistent with the results accounting for $M_B$.

Next, based on the likelihood analysis the star-forming galaxy sample 
has a best-fit model of
\begin{align}
\log\,\bar{W}_r(2796)\, = &\, (1.42\pm0.25)-(1.05\pm0.17) \log\,d \nonumber \\
& + (0.21\pm0.08) \times (M_{\rm star} - M_{\rm star*})
\label{equation:bestfitMstarBlue}
\end{align}
with an intrinsic scatter $\sigma _c=0.27\pm 0.02$ and 
covering fractions of 
$\epsilon_1=0.97\pm 0.03$,
$\epsilon_2=0.64\pm 0.08$,
$\epsilon_3=0.17\pm 0.09$, and
$\epsilon_4=0.02\pm 0.02$.

For quiescent galaxies, the best-fit model with the scaling of stellar mass is
\begin{align}
\log\,\bar{W}_r(2796)\, = &\, (0.01\pm0.36)-(0.24\pm0.21) \log\,d \nonumber \\
& + (0.01\pm0.03) \times (M_{\rm star} - M_{\rm star*})
\label{equation:bestfitMstarRed}
\end{align}
with an intrinsic scatter $\sigma _c=0.38\pm 0.04$ and 
covering fractions of 
$\epsilon_1=0.50\pm 0.16$,
$\epsilon_2=0.26\pm 0.12$,
$\epsilon_3=0.08\pm 0.06$, and
$\epsilon_4=0.09\pm 0.09$.
The results of the likelihood analyses are summarized 
in Table \ref{table:LikelihoodSummary}.

Finally, to assess the significance of the anti-correlation 
without model dependence, we also perform a non-parametric, 
generalized Kendall's $\tau$ test \citep{Feigelson:1985} that 
accounts for the presence of non-detections. The results of Kendall's 
$\tau$ test are presented in Table \ref{table:LikelihoodSummary}.  
We find that the \ewr\ versus $d$ for isolated galaxies deviate from 
a random distribution at more than {an} 11$\sigma$ level.
The \ewr\ is anti-correlated with $d$ at $>10\sigma$ level of 
significance for isolated, star-forming galaxies.  The distribution 
of \ewr\ versus $d$ after accounting for $R_{\rm h}$, $L_{\rm B}$ 
or $M_{\rm star}$ shows similar level of significance.
{For isolated, quiescent galaxy sample, the significance of the 
generalized Kendall test is at $\sim 4\sigma$ level, 
weaker than the significance for star-forming galaxy sample. }
In our likelihood analysis, 
we treat an upper limit as either the QSO sightline intercepts 
\MgII gas but below the detection limit or simply does not 
intercept \MgII gas. The generalized Kendall test considers 
all data to follow a single distribution and therefore sensitive 
upper limits contribute significantly to the anti-correlation 
in the generalized Kendall test compared to the likelihood analysis.

\begin{table*}
\centering
\caption{Summary of Maximum Likelihood Solutions}
\label{table:LikelihoodSummary} 
\centering \resizebox{6.9in}{!}{
\begin{tabular}{lccccccccc}
\hline \hline
Model I & $a_0$ & $a_1$ & $a_2$ & $\sigma_c$ & $\epsilon_1 [0,40]^\dagger$ & $\epsilon_2 [40,100]$ & $\epsilon_3 [100,300]$ & $\epsilon_4 [300,700]$ & ${\rm Kendall's}\,\, \tau$\\
\hline 
full & $0.83\pm0.38$ & $-0.72\pm0.25$ & ...  & $0.32\pm0.04$ & $0.87\pm0.05$ & $0.49\pm0.06$     & $0.19\pm0.12$ & $0.06\pm0.06$  & $-0.46\pm0.04$ \\
blue & $0.94\pm0.30$ & $-0.78\pm0.20$ & ...   & $0.28\pm0.03$ & $0.92\pm0.05$ & $0.56\pm0.09$     & $0.19\pm0.12$ & $0.08\pm0.08$ & $-0.52\pm0.05$\\
red & $-0.08\pm0.34$ & $-0.17\pm0.20$ & ...   & $0.34\pm0.06$ & $0.67\pm0.11$ & $0.41\pm0.10$     & $0.09\pm0.07$ & $0.04\pm0.04$  & $-0.22\pm0.05$\\
\hline 
Model II & $a_0$ & $a_1$ & $a_2$ & $\sigma_c$ & $\epsilon_1 [0,0.2]$ & $\epsilon_2 [0.2,0.5]$ & $\epsilon_3 [0.5,1.1]$ & $\epsilon_4 [1.1,4]$  & ${\rm Kendall's}\,\, \tau$ \\
\hline
Rh-full & $-0.88\pm0.15$ & $-0.84\pm0.20$ & ...  & $0.29\pm0.03$ & $0.85\pm0.06$ & $0.57\pm0.08$ & $0.10\pm0.06$     & $0.07\pm0.07$  & $-0.45\pm0.04$\\
Rh-blue & $-0.98\pm0.17$ & $-1.03\pm0.22$ & ...  & $0.26\pm0.03$ & $0.95\pm0.04$ & $0.67\pm0.08$ & $0.18\pm0.10$     & $0.12\pm0.12$ & $-0.55\pm0.05$\\
Rh-red & $-0.48\pm0.27$ & $-0.12\pm0.24$ & ...  & $0.37\pm0.09$ & $0.53\pm0.18$ & $0.39\pm0.15$ & $0.07\pm0.07$     & $0.05\pm0.05$ & $-0.21\pm0.05$\\
\hline 
Model III & $a_0$ & $a_1$ & $a_2$ & $\sigma_c$ & $\epsilon_1 [0,40]$ & $\epsilon_2 [40,100]$ & $\epsilon_3 [100,300]$ & $\epsilon_4 [300,700]$  & ${\rm Kendall's}\,\, \tau$ \\
\hline
MB-full & $1.22\pm0.25$ & $-0.94\pm0.16$ & $-0.09\pm0.03$  & $0.28\pm0.03$ & $0.95\pm0.04$ & $0.58\pm0.09$ & $0.11\pm0.07$     & $0.04\pm0.04$  & $-0.47\pm0.04$\\
MB-blue & $1.57\pm0.28$ & $-1.14\pm0.18$ & $-0.12\pm0.02$  & $0.25\pm0.03$ & $0.97\pm0.02$ & $0.60\pm0.08$ & $0.18\pm0.11$     & $0.05\pm0.05$ & $-0.56\pm0.05$\\
MB-red & $0.09\pm0.33$ & $-0.27\pm0.20$ & $-0.02\pm0.05$  & $0.34\pm0.06$ & $0.54\pm0.17$ & $0.37\pm0.17$ & $0.07\pm0.07$     & $0.03\pm0.03$ & $-0.23\pm0.05$\\
\hline
Model IV& $a_0$ & $a_1$ & $a_2$ & $\sigma_c$ & $\epsilon_1 [0,40]$ & $\epsilon_2 [40,100]$ & $\epsilon_3 [100,300]$ & $\epsilon_4 [300,700]$  & ${\rm Kendall's}\,\, \tau$ \\
\hline
Mstar-full & $1.35\pm0.25$ & $-1.05\pm0.17$ & $0.21\pm0.08$  & $0.28\pm0.03$ & $0.85\pm0.06$ & $0.53\pm0.07$ & $0.10\pm0.07$     & $0.07\pm0.07$ & $-0.46\pm0.04$ \\
Mstar-blue & $1.42\pm0.22$ & $-1.07\pm0.14$ & $0.28\pm0.06$  & $0.27\pm0.02$ & $0.97\pm0.03$ & $0.64\pm0.08$ & $0.17\pm0.09$     & $0.02\pm0.02$ & $-0.56\pm0.05$ \\
Mstar-red & $0.01\pm0.36$ & $-0.24\pm0.21$ & $0.01\pm0.03$  & $0.38\pm0.04$ & $0.50\pm0.16$ & $0.26\pm0.12$ & $0.08\pm0.06$     & $0.09\pm0.09$ & $-0.22\pm0.05$\\
\hline
\multicolumn{10}{l}{$^\dagger \epsilon_i\,[d_1,d_2]$ represents the mean covering fraction within $[d_1,d_2]$ in $d$ or normalized $d$.}
\end{tabular}
}

\end{table*}

\begin{figure*}
	\centering
	\includegraphics[scale=0.75]{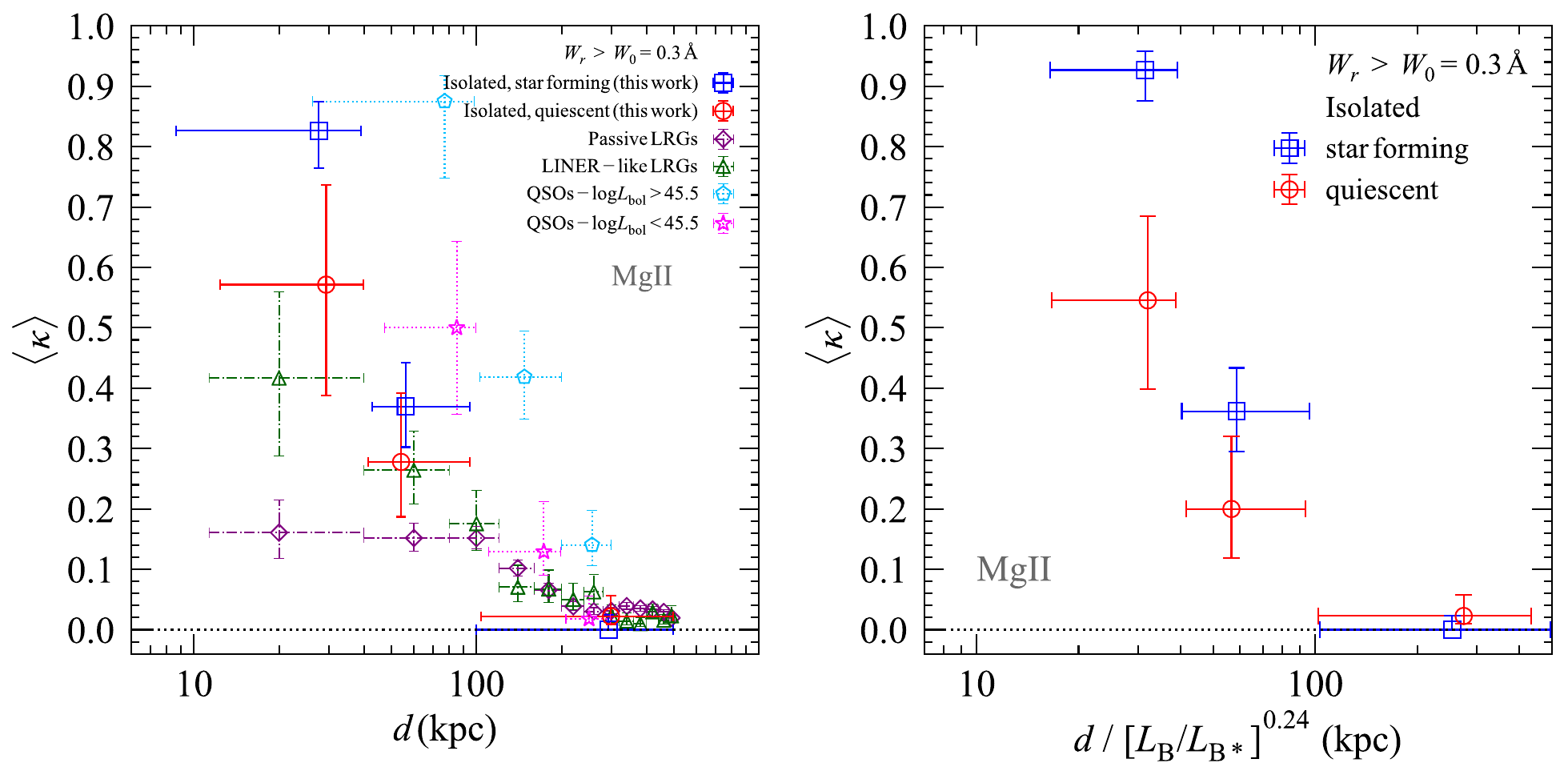}
	\caption{Mean covering fraction of \MgII absorbers $\langle \kappa \rangle$ 
	versus projected distance $d$ ({\it left} panel) and versus 
	$d$ accounting for the galaxy B-band luminosity scaling 
	relation ({\it right} panel).
		We show the gas covering fraction of isolated, 
		star-forming galaxies as blue solid squares, and that of isolated, 
		quiescent galaxies as red solid circles. 
		For comparison, we also include covering fraction measurements 
		from the SDSS LINER-like and passive luminous 
		red galaxy (LRG) samples of \citet{Huang:2016}
	in green dash-dotted triangles and purple dash-dotted 
	diamonds, respectively. We also show the covering fraction 
	for luminous quasars ($\log L_{\rm bol}/{\rm erg\,s^{-1}}>45.5$) in
	light blue dotted pentagons and low-luminosity quasars 
	($\log L_{\rm bol}/{\rm erg\,s^{-1}}<45.5$) in magenta 
	dotted stars \citep{Johnson:2015b}.
	The gas covering fraction is computed for a detection 
	threshold of $W_0=0.3\,\rm \AA$.
	Error bars represent the 68$\%$ confidence interval.
	}
	\label{figure:kappa1}
\end{figure*}

\section{Discussion}
\label{section:discussion}
We have established a spectroscopic sample of 211 isolated 
galaxies and 43 non-isolated galaxies with constraints on 
\MgII absorption from background quasars at projected 
distances of $d<500$ kpc. We characterized the cool gas 
contents of galaxy host halos as a function of projected distance.
We performed likelihood analysis for the isolated galaxies to 
study the dependence of \MgII gas on galaxy projected distance $d$.
Here, we present the observed mean covering fraction ($\kappa$) 
of MgII absorbing gas and examine how the incidence of cool gas 
varies with galaxy properties (i.e. $B$-band magnitude, stellar 
mass and \Halpha\ equivalent width).
We discuss the kinematics, physical conditions and azimuthal 
dependence of these MgII absorbers. 
Finally, we discuss the properties of gaseous halos around 
different host galaxies and compare our results with previous studies.

\subsection{Covering fraction of \MgII absorbers}
In the previous section, we show
that while \MgII absorption equivalent widths of individual 
absorbers decrease with increasing distance for star-forming 
galaxies, no clear trends are seen for quiescent galaxies.  
On the other hand, the best-fit results of the likelihood analysis
suggest a strong anti-correlation between covering fraction 
of \MgII gas ($\epsilon$) and projected distance 
for both samples (Table \ref{table:LikelihoodSummary}).  
Take model I for example, the best-fit \MgII gas covering 
fraction for star-forming galaxies
at $d<40\,\rm kpc$ is $\epsilon_1=0.87\pm0.05$, 
which declines to $\epsilon_2=0.49\pm0.06$ at $d=40-100\,\rm kpc$.
For quiescent galaxies, the covering fraction is 
$\epsilon_1=0.67\pm0.11$ at $d<40\,\rm kpc$, which
subsequently declines to $\epsilon_2=0.41\pm0.10$ at 
$d=40-100\,\rm kpc$. Beyond $d=100\,\rm kpc$, both samples 
show covering fraction consistent with $\epsilon \approx 0$ 
within 2-$\sigma$ level. Similar anti-correlations are also 
reported in the best-fit results of Model II - Model IV, 
regardless of various scaling relations with 
halo radius, $B-$band magnitude and stellar mass.

We note that the \MgII covering fraction ($\epsilon$) derived 
using the likelihood analysis denotes the 
{\it intrinsic} incidence of \MgII gas, which does not 
depend on the absorption strength of \MgII.
{In practice, given the sensitivity of the background QSO spectra,
our survey does not provide much information for
\MgII absorbers weaker than 0.1\AA.}
Here we also report the observed mean covering 
fraction ($\kappa$) of \MgII absorbing gas 
and examine how the incidence of cool gas 
varies with galaxy properties.
Following the prescription in \cite{Chen:2010},
we employ a maximum likelihood analysis to estimate
$\kappa$ and its uncertainties.
The likelihood of detecting an ensemble of galaxies
of which $n$ galaxies with associated \MgII absorbers 
of $W_r \geq W_0$ and $m$
galaxies with no absorbers detected down to a 
sensitive limit of $W_r < W_0$ is
  \begin{equation}
  \begin{split}
  {\mathscr L}(\kappa | W_0) = \langle \kappa \rangle^n \left[1-\langle \kappa \rangle\right]^m
  \end{split}
  \end{equation}
We evaluate $\kappa$ for a detection threshold of 
$W_0=0.3\rm\, \AA$. In the isolated galaxy samples, 
201/211 MagE spectra of corresponding QSOs have
sufficient S/N for detecting \MgII gas with absorption 
strength exceeding ${W}_r(2796)\geq0.3\rm\, \AA$.
We evenly divide each sample into various projected distance
intervals, and obtain best-fit observed covering fractions 
($\kappa$) and associated uncertainties for each
projected distance bin.  The results are shown in the left 
panel of Figure \ref{figure:kappa1}.

For the star-forming galaxy sample, the covering fraction 
of ${W}_r(2796)\geq 0.3\,$\AA\ absorbers
 declines from $\left<\kappa\right>=0.83^{+0.06}_{-0.05}$ at $d<40\,\rm kpc$ to $\left<\kappa\right>=0.37^{+0.07}_{-0.07}$ at 
 $d\approx40-90\,\rm kpc$.  Beyond $d\approx100\,\rm kpc$, 
 the covering fraction declines to $\left<\kappa\right>\approx 0$.
Quiescent galaxies exhibit similar but mildly lower observed 
\MgII covering fraction at inner projected distances ($<90\,\rm kpc$).
At $d<40\,\rm kpc$, $\left<\kappa\right>=0.57^{+0.16}_{-0.18}$ 
for quiescent galaxies, which decreases to 
$\left<\kappa\right>=0.28^{+0.11}_{-0.09}$ at
$d\approx40-90\,\rm kpc$ and subsequently to 
$\left<\kappa\right>=0.02^{+0.03}_{-0.01}$ beyond $d\approx100\,\rm kpc$.
The observed anti-correlations between covering fraction 
and projected distance for both samples are consistent 
with the best-fit results from the likelihood analysis. 
We find that star-forming galaxies seem to have elevated 
incidence of \MgII gas at inner projected distance of 
$d<40\,\rm kpc$ compared to our quiescent galaxy sample.

To take into account the possible correlation between 
covering fraction and galaxy properties,
we also calculate the mean observed covering 
fractions in different $M_{\rm B}$-normalized projected
distance intervals, using the best-fit scaling relation 
of Equation (\ref{equation:bestfitMstarBlue}).
The results are shown in the right panel of Figure \ref{figure:kappa1}.
After applying the $M_{\rm B}$-scaling,
The anti-correlation between covering fraction 
and normalized projected distance for star-forming galaxies is strengthened.
The \MgII gas covering fraction for star-forming 
galaxies is $\left<\kappa\right>=0.93^{+0.03}_{-0.04}$ 
at $d^{\prime}<40\,\rm kpc$ and $\left<\kappa\right>=0.36^{+0.07}_{-0.07}$ 
at $d^{\prime}=40-100\,\rm kpc$,
where $d^{\prime}=d\times10^{0.10(M_B-M_B^*)}$ is the 
$M_{\rm B}$-normalized projected distance.
For quiescent galaxies, we obtain
 $\left<\kappa\right>=0.55^{+0.14}_{-0.15}$ at $d^{\prime}<40\,\rm kpc$ and 
 $\left<\kappa\right>=0.2^{+0.12}_{-0.08}$ at $d^{\prime}=40-100\,\rm kpc$.
 At inner $d^{\prime}\lesssim40\,\rm kpc$, 
 the incidence of \MgII gas seems to be suppressed
 in the quiescent galaxy sample compared to that of 
 star-forming galaxies at $\sim$ 3-$\sigma$ level.
The fact that the difference in covering fraction between 
the two samples is more evident after the 
$M_{\rm B}$-scaling highlights that 
there is an unambiguous connection between the 
\MgII gas and host galaxy properties.

Indeed, the high \MgII gas covering fractions of both 
star-forming and quiescent galaxy samples at $d<90$ kpc
are in stark contrast to what we have seen in luminous red 
galaxies \citep[LRGs;][]{Huang:2016}, which we also show 
in the left panel of Figure \ref{figure:kappa1} for comparison.
In \cite{Huang:2016}, we utilized $\sim 38000$ LRG-QSO 
pairs in SDSS DR12 and divided the LRGs into passive 
(purple diamonds) and LINER-like subsamples (green triangles)
according to whether they exhibit \OII-emission features.
We reported a constant \MgII gas covering fraction of 
merely $\approx 15\%$ at $d\lesssim$ 120 kpc 
for passive LRGs, which comprises the majority of the 
LRG sample ($\sim 90\%$). The LINER-like LRGs have a 
slightly elevated covering fraction of $\approx 40\%$ at $d<40\,\rm kpc$,
which declines to a similar level of $\approx 15\%$ at 
$d\approx\rm100\,kpc$. We also display the measurement 
of quasar host halos from \cite{Johnson:2015b} for comparison.  
The luminous quasars of bolometric luminosity 
$\log{\it L}_{bol}/({\rm erg}\,{\rm s}^{-1})>45.5$ at 
$\langle z \rangle \approx 1$ are shown in light blue pentagons 
and low-luminosity ones of $\log{\it L}_{bol}/({\rm erg}\,{\rm s}^{-1})<45.5$ 
at $\langle z \rangle \approx 0.75$ are shown in magenta stars.  
Both the luminous and low-luminosity quasars exhibit substantial 
amount of \MgII absorbing gas ($\kappa \approx 0.42$ and 
$\kappa \approx 0.13$ respectively) at $d=100-200$ kpc, 
and elevated \MgII covering fraction of $\kappa \approx 0.9$ 
and $\kappa \approx 0.5$ within 100 kpc compared to our 
isolated star-forming and quiescent galaxy samples.

We note that LRGs are luminous and massive 
($M_{\rm star}\sim10^{11.4}\,M_{\odot}$), residing in halos of
$M_{\rm halo} \gtrsim 10^{13}\,M_{\odot}$ 
\citep[e.g.,][]{Zhu:2014,Huang:2016}. 
The galaxies in our samples are primarily fainter, 
spanning a broad range of stellar mass from $10^{8.6}-10^{11.5}$, 
with a median stellar mass $M_{\rm star}=10^{10.5}$.
Using our large galaxy samples with sensitive limits 
of \MgII absorption, we can examine how the observed 
mean \MgII covering fraction depends on the stellar mass.
To perform a representative comparison across a broad 
range of stellar mass, the mean covering fraction is calculated 
based on observations of \MgII absorbing
gas within a fiducial halo gas radius, 
$R_{\rm gas}$ \citep{Kacprzak:2008,Chen:2010}.
In \cite{Chen:2010}, we used a sample of 71 isolated 
galaxies from this survey to show that the extent of 
\MgII absorbing haloes is well described
by an isothermal density profile with a boundary at $d=R_{\rm gas}$, 
that scales with galaxy $B$-band luminosity according to
$R_{\rm gas} \approx 107 \times (L_B/L_B^*)^{0.35}\, \rm kpc$.
We equally divide our star-forming galaxies into two mass intervals, 
with median stellar masses of $10^{9.7}$ and $10^{10.5} \,M_{\odot}$.
The corresponding $R_{\rm gas}$ is respectively around 
$ 70\,\rm kpc$ and $ 100\,\rm kpc$. For LRGs with a mean 
luminosity of $\approx 3.6L_*$, we infer $R_{\rm gas} \approx 206\,\rm kpc$.

\begin{figure*}
	\centering
	\includegraphics[scale=0.7]{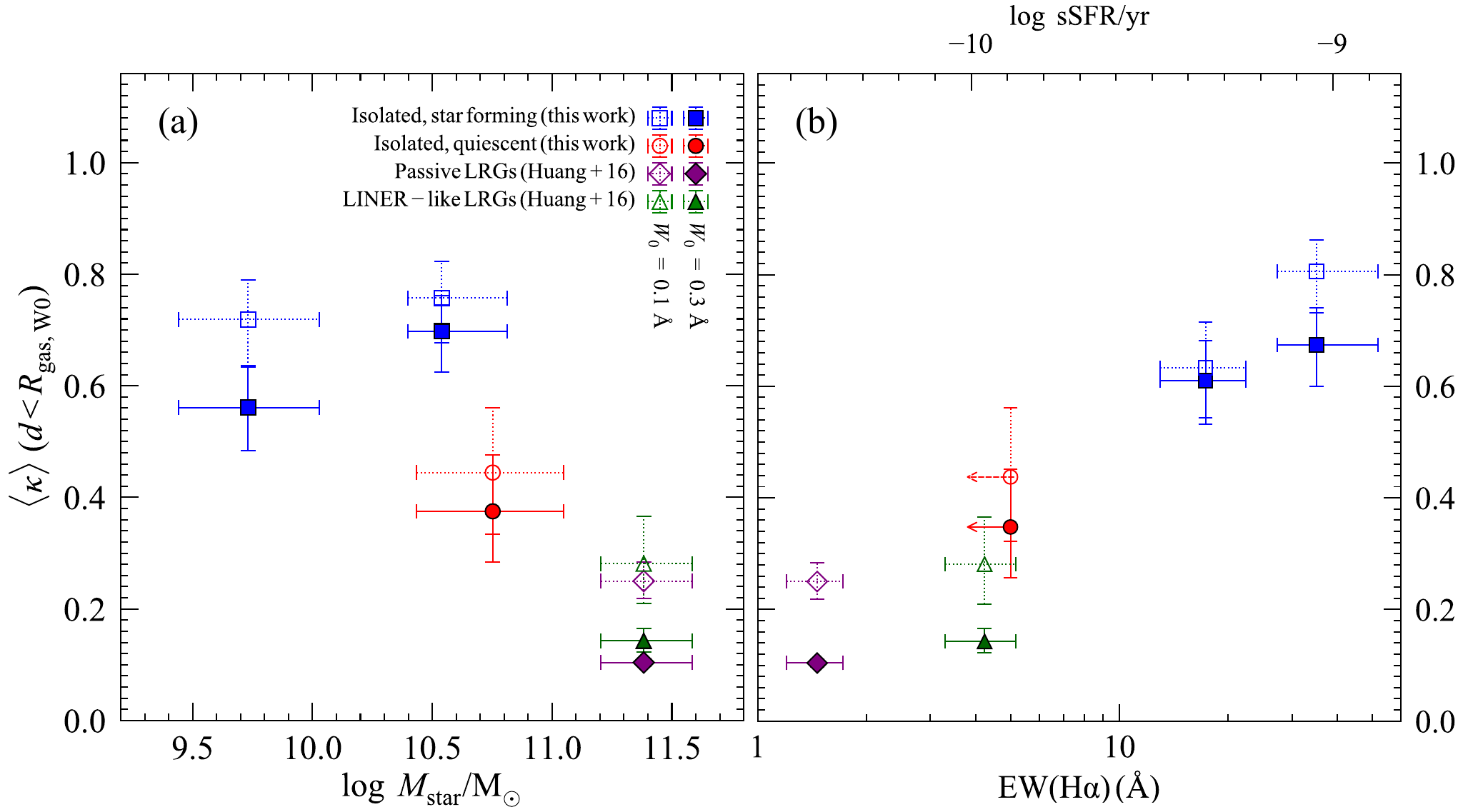}
	\caption[]{(a) Mass dependence of the mean covering 
	fraction of \MgII absorbing gas.
	Isolated, star-forming galaxies are shown in blue squares, 
	and quiescent galaxies are shown in red circles.
	Constraints from passive luminous red galaxies (LRGs) 
	and \OII-emitting LRGs
	are included as purple diamonds and green triangles \citep{Huang:2016}.
	Solid points indicate absorbers of $W_0=0.3\,\rm \AA$, 
	and open points indicate absorbers of $W_0=0.1\,\rm \AA$.
	The horizontal bars represent 68\% range of stellar mass for 
	galaxies included in each bin, and the vertical error
	bars show the 68\% confidence interval. 
	(b) \Halpha\ equivalent width versus the mean covering 
	fraction of \MgII absorbing gas. The symbols are the same as (a). 
	The inferred specific star-formation rate (sSFR) on the 
	top axis is derived from Section \ref{section:galproperties}.
	Note that the \Halpha\ emission for some passive galaxies
	are likely due to the underlying AGNs or LINERs, 
	and therefore the inferred
	sSFR based on the observed \Halpha\ emission 
	only represents an upper limit.}
	\label{figure:LRGs}
\end{figure*}

We present in the {\it left} panel of Figure \ref{figure:LRGs} the 
mean gas covering fraction $\langle \kappa \rangle$ within $R_{\rm gas}$ 
for our star-forming and quiescent galaxy samples,
and the LRG samples in \cite{Huang:2016}.
We show $\langle \kappa \rangle$ for absorption equivalent 
width thresholds $W_{0}=0.3\rm \,\AA$ (solid symbols) and
$W_{0}=0.1\rm \,\AA$ (open symbols).
A number of interesting features are revealed in this plot.

At the $W_{0}=0.3\rm \,\AA$ threshold, firstly the star-forming 
galaxies display a slightly elevated gas covering fraction 
from $\langle \kappa \rangle=0.56^{+0.07}_{-0.08}$ at 
$M_{\rm star}\approx10^{9.7}\,M_{\odot}$
to $\langle \kappa \rangle=0.70^{+0.06}_{-0.07}$ at 
$M_{\rm star}\approx10^{10.5}\,M_{\odot}$.
Second, there is an apparent difference in covering fraction between 
the star-forming and quiescent galaxy samples for $L_*$ galaxies.
The covering fraction for quiescent galaxies 
is $\langle \kappa \rangle=0.30^{+0.10}_{-0.09}$ at 
$M_{\rm star}\approx10^{10.8}\,M_{\odot}$, merely 
$\approx40\%$ of what we observe in star-forming galaxy halos.
Finally, a sharp decline of \MgII gas covering fraction
is revealed from even quiescent $L_*$ galaxies to massive LRGs.
Specifically,  $\langle \kappa \rangle$ declines from 
$30-70\%$ around $L_*$ galaxies
to $\approx10-15\%$ around massive LRGs.
The strong mass dependence of \MgII covering fraction
qualitatively agrees with the expectation from the observed 
clustering of \MgII absorbers, where the \MgII covering fraction 
peaks at $M_{\rm halo}\sim10^{12}\,M_{\odot}$ 
($\sim L_*$ galaxies) and rapidly falls off at smaller and higher masses \citep{Tinker:2008,Tinker:2010}.

In the {\it right} panel of Figure \ref{figure:LRGs} we present 
the mean gas covering fraction $\langle \kappa \rangle$ 
within $R_{\rm gas}$ as a function of \Halpha\ equivalent width
for our MagE galaxies and the LRG samples in \cite{Huang:2016}.  
The symbols are the same as in the left panel.  
We find that there is a positive correlation between 
$\langle \kappa \rangle$ and \Halpha\ equivalent width 
(or star formation rate).  We note that although LINER-like LRGs 
(green triangles) have significantly lower $\langle \kappa \rangle$ 
compared to the MagE quiescent galaxy sample, in \cite{Huang:2016} 
we find that their \Halpha\ is likely contributed by post-asymptotic 
giant branch (post-AGB) stars.  Therefore, the star-formation rates 
of LRG samples represent upper limits.

\begin{figure*}
	\centering
	\includegraphics[scale=0.85]{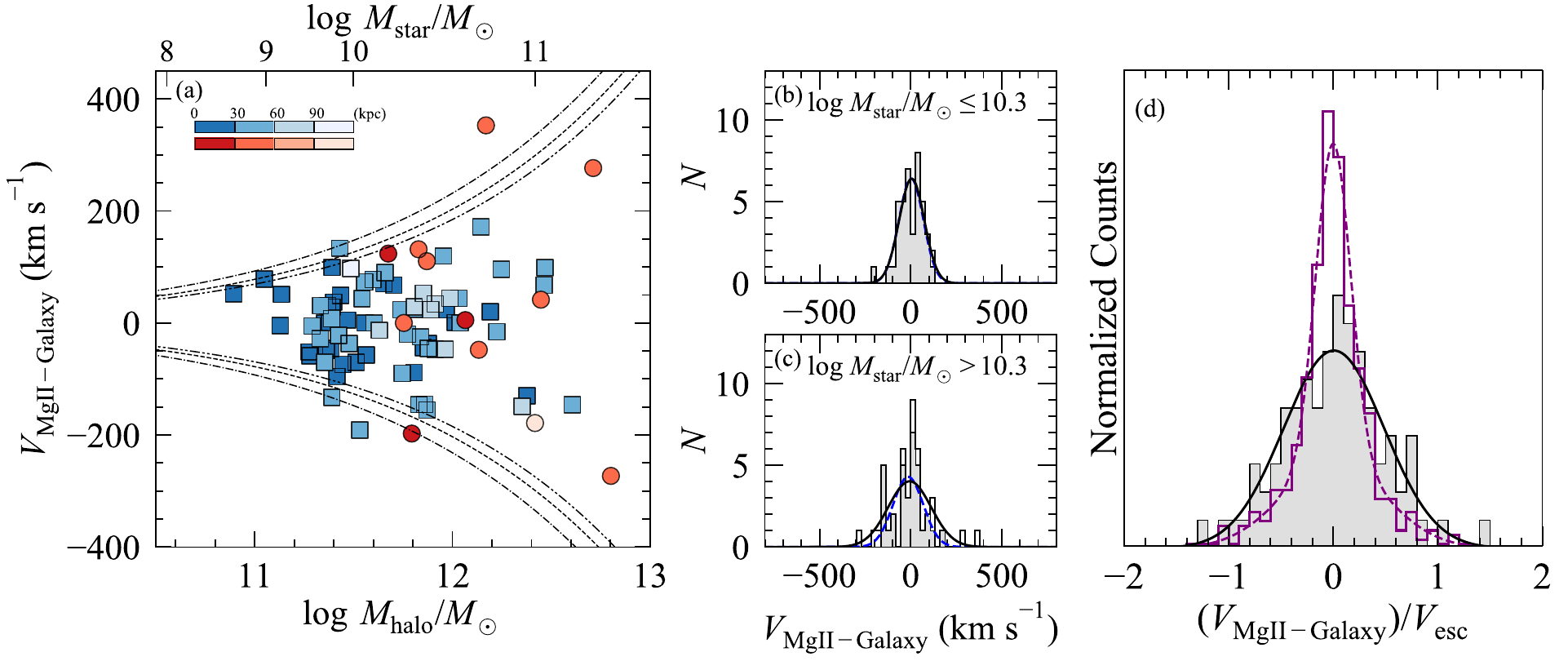}
	\caption{{\it Left:} Relative \MgII absorption velocities with respect
	to the galaxy systematic redshifts as a function of inferred 
	dark matter halo mass 
	for isolated, star-forming (blue squares) and quiescent 
	(red circles) galaxies. 
	The darkness of the blue/red symbols indicates galaxies 
	at various projected distances
	according to the color bars in the {\it top left} corner.
	The three pairs of dashed curves from outside to inside 
	mark the line-of-sight projected halo escape
	velocities at distance $r$=30, 60, and 90 kpc, assuming 
	an Navarro-Frenk-White (NFW) profile of dark matter halos.
	 {\it Middle:} The relative velocity distributions of \MgII 
	 absorbers for low-mass ($M_{\rm star}\leq10^{10.3}\,M_{\odot}$; {\it top middle} panel)
	  and high-mass ($M_{\rm star}>10^{10.3}\,M_{\odot}$; 
	  {\it bottom middle} panel) galaxies.
	 The filled and open histograms represent respectively 
	 the number counts for isolated full and star-forming galaxy samples.
	Adopting a Gaussian profile to characterize the velocity
	distribution of \MgII gas around low-mass (high-mass) galaxies 
	leads to a Gaussian profile centered at 
	$\langle v_{{\rm MgII-Galaxy}}\rangle= 6 (-5)$ \kms\ and
	$\sigma$ = 69 (117) \kms (black solid curves).
	Considering only star-forming galaxies for the 
	low-mass (high-mass) galaxy sample,
	we find a mean and dispersion of 
	$\langle v_{{\rm MgII-Galaxy}}\rangle= 3 (-13) $ \kms\ and $\sigma$ = 67 (83) \kms (blue dashed curves).
	{\it Right:} 	The relative velocity distributions of \MgII absorbers 
	normalized by the line-of-sight projected halo 
	escape velocities ($V_{\rm esc}$) at the projected distance
	for isolated galaxies (filled histogram; this work) and 
	passive LRGs \citep[purple open histogram;][]{Huang:2016}.  
	\MgII-absorbing gas around isolated galaxies can 
	be characterized by a
	single Gaussian distribution centered at 
	$\langle v_{{\rm MgII-Galaxy}}\rangle / v_{\rm esc}= 0.01 $ and $\sigma$ = 0.48 (black curve). 
	Applying a double Gaussian profile similar to 
	\citet{Huang:2016},
	\MgII-absorbing gas around passive LRGs can be 
	characterized by a narrow component
	centered at $\langle v_{{\rm MgII-Galaxy}}\rangle / v_{\rm esc}= 0.00 $ 
	and $\sigma$ = 0.17
	and a broad component centered at 
	$\langle v_{{\rm MgII-Galaxy}}\rangle / v_{\rm esc}= 0.00$ and $\sigma$ = 0.50
	(purple dashed curve).  
	}
	\label{figure:vdiff_isosf}
\end{figure*}

\subsection{Kinematics}
\label{Kinematics}
The line-of-sight velocity dispersion of \MgII absorbers provides 
important insights into the underlying motion and physical nature of cool
clumps within host halos. 
The left panel of Figure \ref{figure:vdiff_isosf} displays 
the relative velocity between MgII absorbers and their host 
galaxies as a function of inferred dark matter halo mass.
The dashed curves mark the projected escape velocities at 
$r$=30, 60, 90 kpc (from outside to inside) with respect to mass. 
No significant trends are observed between the relative velocity 
and the projected distance between QSO-galaxy pairs. 
Below projected distance $d<40\,{\rm kpc}$, we find the median 
relative velocity difference to be $\langle |\Delta v | \rangle=48$ \kms.  
At $d\geq40\,{\rm kpc}$, the relative velocity difference is 
$\langle | \Delta v | \rangle=47$ \kms.
At similar stellar masses, quiescent galaxies seem to have higher 
relative velocity offsets from the galaxy systematic redshifts 
compared to that of star-forming galaxies.
The majority of the detected \MgII absorbing gas is found at 
velocities below the expected projected escape velocities,
indicating that these \MgII gas complexes are likely to be 
gravitationally bound.

Using the large sample of galaxy and absorber pairs in 
the \MAGMAGMAG\ Halo Project,
 we are able to constrain how the ensemble average of the velocity
distribution of absorbing gas changes with their host galaxy properties.
We divide isolated galaxy sample into 
low-mass ($\log \langle M_{\rm star}/\msun\rangle \approx{9.7}$) 
and high-mass ($\log \langle M_{\rm star}/\msun \rangle\approx {10.6}$) 
galaxies to investigate the 
correlation between mass and velocity dispersion in the 
middle panels of Figure \ref{figure:vdiff_isosf}.
The velocity distribution is 
well characterized by a Gaussian of dispersion $\sigma=69\,\rm km\,s^{-1}$ 
around low-mass galaxies and
 $\sigma=117\,\rm km\,s^{-1}$ for high-mass galaxies.
Quiescent galaxies seem to have higher relative velocities 
than that of star-forming galaxies. If we consider only star-forming 
galaxies, we find that the relative velocity dispersion of low-mass 
galaxies is $\sigma=67\,\rm km\,s^{-1}$,  and is 
$\sigma=83\,\rm km\,s^{-1}$ for high-mass galaxies.  
The velocity dispersion of the high-mass sample is $\approx20\%$ 
more elevated than the low-mass sample.  

We use the bootstrap method to estimate the 68\% confidence levels 
of the relative velocity dispersion.  For isolated star-forming galaxies, 
the low-mass sample gives $\sigma=58-73\,\rm km\,s^{-1}$, while the 
high-mass sample gives $\sigma=73-90\,\rm km\,s^{-1}$.
There seems to be a positive correlation between the velocity 
dispersion of \MgII absorbing gas and mass of associated host halos.
Assuming a NFW profile \citep{Navarro+97} with halo concentration 
of $c_{\rm h}=10$, we can calculate the expected line-of-sight velocity 
dispersion for virialized motion within $d=100\,{\rm kpc}$,
beyond which no \MgII gas is detected in the isolated star-forming 
galaxy sample. The expected line-of-sight velocity dispersion for 
low-mass and high-mass galaxies
are respectively $\sigma=52\,\rm km\,s^{-1}$ and 
$\sigma=86\,\rm km\,s^{-1}$,
comparable to the observed velocity dispersion.
The recent work on \MgII absorbing gas around 50 
star-forming galaxies at $z\approx0.2$ \citep{Martin:2019} 
also shows a consistent result.  With a median stellar mass 
of $M_{\rm star}=10^{10}\, M\rm_{\odot}$, the sample has a 
relative velocity dispersion of $\sigma=42-54\,\rm km\,s^{-1}$, 
comparable to the expected line-of-sight velocity dispersion 
of $\sigma=56\,\rm km\,s^{-1}$.  

Our result is in stark contrast to the \MgII gas in LRG halos, 
where the line-of-sight velocity dispersion is merely 60 percent 
of what is expected from virial motion 
\citep{Huang:2016,Zhu:2014,Zahedy:2019,Afruni:2019}.
In the right panel of Figure \ref{figure:vdiff_isosf},
we display the relative velocity distributions of \MgII absorbers
divided by the line-of-sight projected halo escape velocity for 
our isolated galaxies (filled histogram) and passive LRGs in 
\citet{Huang:2016} (open histogram).
It is clear that while for both samples the majority of the 
detected \MgII absorbing gas
is at velocities well within the expected projected escape velocities,
\MgII gas found around passive LRGs have suppressed velocity dispersion
compared to our isolated galaxies.
{Since the Kolmogorov-Smirnov (K-S) test tends to be most sensitive
around the median values of distributions and less sensitive to the tails
of distributions, we use Anderson-Darling (A-D) test \citep{Anderson:1952}, 
which provide increased sensitivity on the shapes of distributions.
The A-D test shows that the probability of the two
velocity dispersions of \MgII absorbers to be drawn
from the same distribution is $P\sim1.4$\%.}

The comparable velocities between observation and the 
expectation from virial motion support the physical formalism 
for a two-phase CGM, where QSO absorption systems in the 
vicinity of galaxies originate in cool clumps which are in thermal 
pressure equilibrium with the hot halo \citep[][]{Mo:1996}.
The positive correlation between the mean gas covering 
fraction $\langle \kappa \rangle$ and \Halpha\ equivalent width (sSFR)
in the {\it right} panel of Figure \ref{figure:LRGs} 
also hints on the possibility that the cool clumps may be able 
to survive and reach the central galaxy.

If clouds are sufficiently massive, they can travel at a characteristic 
speed equal to the halo velocity as the clouds move through 
host gas halo.  Following \cite{Maller:2004}, we are able to place 
a lower limit on the cloud mass using their Equation (40),
\begin{equation}
m_{\rm cl} \approx 5.1\times 10^4 M_{\odot}\,T_6^{-3/8} (\Lambda_z t_8)^{1/2}
\label{equation:cloudmass}
\end{equation}
where $T_6=T/10^6 \rm\, K$ is the temperature of hot halo gas,
$\Lambda_z$ is a cooling parameter that depends on the gas metallicity
and $t_8=t_f/8$ Gyr is the halo formation timescale.
For our low-mass and high-mass galaxy samples of 
$\log \langle M_{\rm star}/\msun\rangle \approx {9.7}$ and 10.6, 
$T\sim5\times10^5\,\rm K$ and $T\sim 10^6 $K assuming an 
isothermal gas, and $t_f\sim9\,\rm Gyr$ according to N-body 
simulations \citep{Wechsler:2002}.  We find the lower limits of 
cloud mass are $m_{\rm cl} = (0.7, 1.5) \times 10^5\, \rm M_{\odot}$ 
for low-mass galaxies and 
$m_{\rm cl} = (0.5, 1.1) \times 10^5\, \rm M_{\odot}$ for high-mass 
galaxies assuming (0.1, 1.0) solar metallicity.
The cloud mass is qualitatively consistent with the initial 
cloud mass of $\approx 10^{4-5}\, \rm M_{\odot}$ 
at the virial radius in \cite{Afruni:2019}.
Unlike massive LRG halos, the inflow accretion of 
gas clouds from external parts of $L_*$ galaxy halos does 
not suffer from severe deceleration by the hot gas 
drag force, and therefore remain massive in the internal regions.

We estimate the free-fall time ($\tau_{\rm ff}$) for the clouds 
to reach the center to be $\tau_{\rm ff}\sim0.29 $ Gyr for 
our \MgII absorbers at projected distance 
$\langle d\rangle _{\rm med}\approx 33$ kpc and 
$\langle z\rangle _{\rm med}\approx0.23$.
This timescale is about an order of magnitude smaller 
than the characteristic timescale for the clouds being 
evaporated by conduction from the surrounding hot gas.
If we assume it takes a free-fall time ($\tau_{\rm ff}$) for 
the clouds to reach the center of the host halo, we can 
infer a cool gas accretion rate of 
\begin{align}
\dot{M} = 1.33 \left( \frac{f_{\rm cl}}{0.1}\right) \left( \frac{m_{\rm cl}}{10^5 M_\odot}\right) \left( \frac{r_{\rm c}}{1\, \rm kpc}\right) \,\, M_{\odot}\,\rm yr^{-1}
\end{align}
where $f_{\rm cl}$ is the volume filling factor of \MgII gas 
within $\approx$ 33 kpc, $m_{\rm cl}$ is the cloud mass, 
and $r_{\rm c}$ is the mean clump size.  Assuming $f_{\rm cl}=0.1$ 
and $r_{\rm c}=1$ kpc, we can obtain a lower limit of cool 
gas accretion rate of $\dot{M} = (0.9,2.0)\,M_{\odot}\,\rm yr^{-1}$ 
for low-mass galaxies and $\dot{M} = (0.7,1.5)\,M_{\odot}\,\rm yr^{-1}$ 
for high-mass galaxies assuming (0.1, 1.0) solar metallicity.  
The estimated accretion rate is similar to the star
formation rate of the Milky Way of $\sim 2\,M_{\odot}\,\rm yr^{-1}$ 
\citep{Chomiuk:2011}. 
Note that given the lower limit of cloud mass we obtain from the Equation (\ref{equation:cloudmass}), 
the characteristic timescale ($\tau_{\rm evap}$) for 
clouds being evaporated by conduction from the surrounding 
hot gas is about $\tau_{\rm evap}\gtrsim 10 \tau_{\rm ff}$.  
Therefore, the cool clumps are likely to reach the center of the host halo.

\begin{figure*}
	\centering
	\includegraphics[scale=0.58]{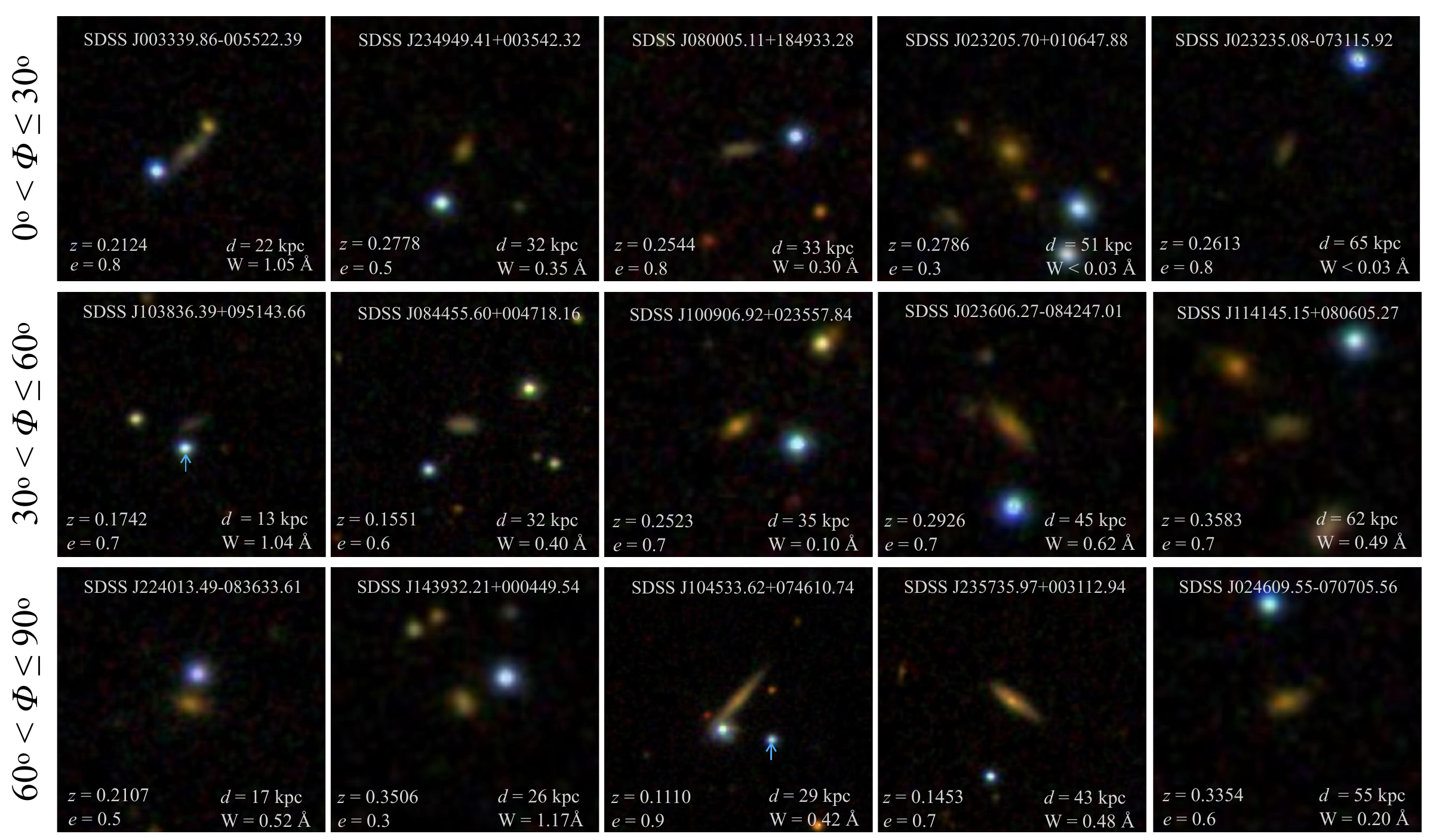}
	\caption{Images of 15 randomly selected galaxy-QSO pairs 
	in our sample to demonstrate that the accuracy of the 
	azimuthal angle ($\Phi$) measurements from SDSS
	is sufficient for the subsequent $\Phi$-dependence study.  
	Each panel is 150 kpc on a side at the galaxy redshift.
	Each galaxy is
	located at the center and the QSO appears as a blue 
	compact source near the galaxy.
	We mark QSOs with blue arrows in some panels for clarification.
	The projected distance ($d$) and the strength of \MgII absorbers 
	(W; measurement or 2$\sigma$ upper limits) of 
	each galaxy are shown in the lower-right corner, 
	and the galaxy redshift ($z$) and the ellipticity ($e$) are 
	shown in the lower-left corner.
	The top, middle and bottom rows show examples of galaxies 
	with azimuthal angle $\Phi$
	in the range of $0\degree < \Phi \leq 30\degree$, $30\degree < \Phi \leq 60\degree$, 
	and
	$60\degree < \Phi \leq 90\degree$, respectively. 
	The images in each row are displayed from left to right in 
	increasing projected distance $d$.}  
	\label{figure:imagePA}
\end{figure*}

\begin{figure*}
	\centering
	\includegraphics[scale=0.6]{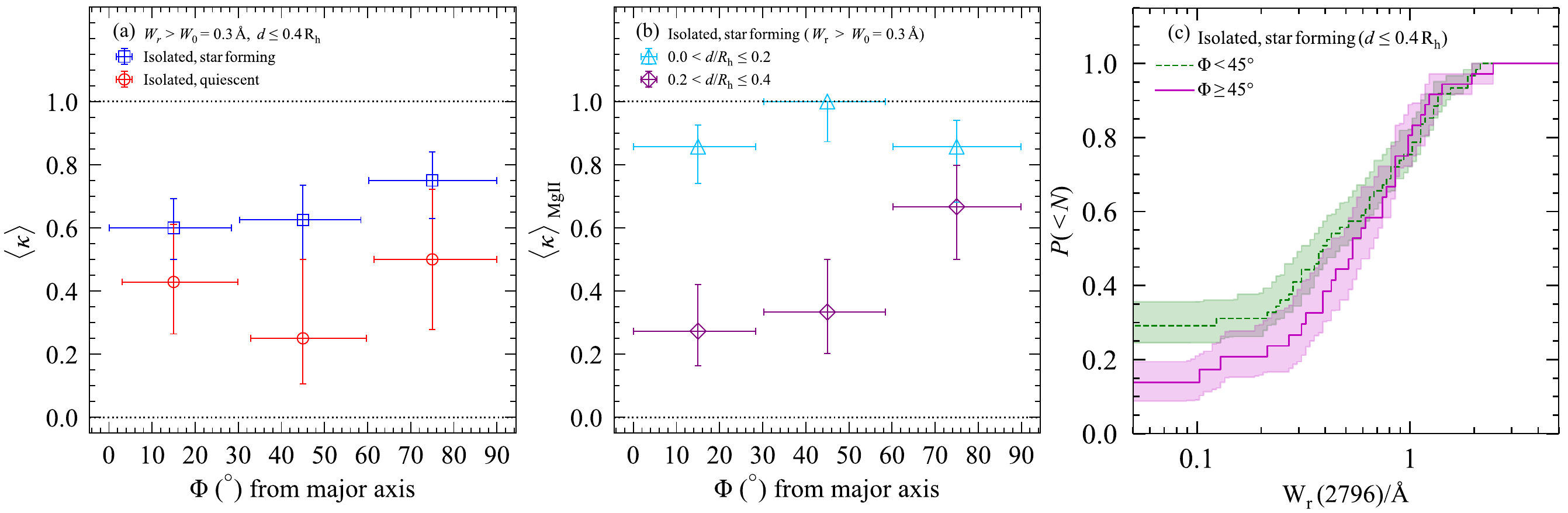}
	\caption{(a) Dependence of $\left<\kappa\right>$
	on the azimuthal angle for isolated, star-forming galaxies (blue squares)
	and quiescent galaxies (red circles) at projected distance $d<0.4\,R_{\rm h}$.
 	(b) Dependence of $\left<\kappa\right>$
	on the azimuthal angle in different projected distance intervals 
	for isolated, star-forming galaxies.
	Triangles represent galaxies within projected distance 
	$d$ of $d\leq0.2\,R_{\rm h}$
	from a background QSO sightline, and diamonds represent 
	$0.2\,R_{\rm h} < d \leq 0.4\,R_{\rm h}$
	The horizontal error bars show the full range of galaxy-QSO 
	projected distances within
	 each bin and vertical error bars represent the 68\% confidence interval.
	 (c) Cumulative fraction ($P$) of isolated, 
	 star-forming galaxies at $d<0.4\,R_{\rm h}$
	 with the rest-frame equivalent width of \MgII no greater than $\ewr$/\AA.
	 The galaxies with azimuthal angle $\Phi<45\degree$ are 
	 shown in dashed green line
	 and galaxies with $\Phi \geq 45\degree$ as magenta solid line.
	 The cumulative distributions are estimated using Kaplan-Meier estimator.
	 The shaded bands represent 68\% confidence intervals 
	 including uncertainties due to sample variance and \ewr\ measurement errors
	 with combined bootstrapping and Monte-Carlo resampling.
	 }  
	\label{figure:PA}
\end{figure*}

\subsection{Angular Distribution of \MgII Absorption Relative to Galaxy Major Axis}
\label{AngularDistribution}
To examine the azimuthal dependence of $\left<\kappa\right>$, 
we calculate the azimuthal angle ($\Phi$)
of each galaxy following the same procedure in \cite{Huang:2016}.
Briefly, $\Phi$ is calculated using the position angle (PA) 
and ellipticity ($e$) measurements from the SDSS database.
The azimuthal angle ($\Phi$) is defined
as the angle from the galaxy's major axis to the line 
connecting the center of the galaxy to the location of the QSO. 
A QSO sightline that occurs along the major
axis of the galaxy has $\Phi=0\degree$ and one that 
occurs along the minor axis of the galaxy has $\Phi=90\degree$.

To ensure a high confidence in the following azimuthal 
dependence investigation, we restrict our sample to those galaxies 
with measured ellipticities $e>0.2$ and consistent
measurements of PA in both SDSS $r$ and $i$ bands.  
SDSS $r$ and $i$ bands are
the most sensitive bandpasses for measuring the surface 
brightness profiles and $\Phi$ for galaxies at $z\approx0.2$ in our sample.
A total of 128 out of 201 galaxies meet this criterion.
We then divide these galaxies into three bins of $\Phi$, where
the bin size is chosen to be larger than the typical 
uncertainty in our data ($\Delta \Phi <10\degree$).
In Figure \ref{figure:imagePA}, we display 15 galaxy-QSO pairs to 
illustrate typical cases
with azimuthal angles falling in the three bins:
$0\degree<\Phi\leq30\degree$ (top), $30\degree<\Phi\leq60\degree$ (middle),
and $60\degree<\Phi\leq90\degree$ (bottom) that 
go into the calculations in Figure \ref{figure:PA}.
From visual inspection of Figure \ref{figure:imagePA}, 
it is clear that the 
measurements of $\Phi$ for the restricted sample 
are sufficiently adequate
for the adopted bin size $\Delta \Phi = 30 \degree$.

We present in the {\it left} panel of Figure \ref{figure:PA}
the $\left<\kappa\right>$
as a function of $\Phi$ for our star-forming and quiescent 
galaxy samples. As we only have one detected \MgII absorber 
for the isolated galaxy sample beyond 0.4$R_{\rm h}$, 
we limit our investigation to $d\leq0.4\,R_{\rm h}$.
We find no strong dependence of $\left<\kappa\right>$
on $\Phi$ for either star-forming or quiescent galaxies.
While we find a modest enhancement ($\approx$ 10 
percent at 1$\sigma$ level) of \MgII absorption closer to the
minor axis ($\Phi \gtrsim 60 \degree$) of star-forming galaxies,
no azimuthal angle preference is found for quiescent galaxies. 
In the {\it middle} panel of Figure \ref{figure:PA}, we further 
divide star-forming galaxies into two projected distance bins.
We find no azimuthal angle dependence of $\left<\kappa\right>$ 
for star-forming galaxies at small projected distance
$0<d\leq0.2\,R_{\rm h}$, and a mild elevated
covering fraction along the minor axis ($\sim 1\sigma$ level) at 
$0.2<d\leq0.4\,R_{\rm h}$. To further investigate the strength of 
\MgII absorbers around galaxies closer to the major ($\Phi<45\degree$) 
and minor ($\Phi \geq 45 \degree$) axis,
we use the Kaplan-Meier estimator \citep{Feigelson:1985} to derive
the median rest-frame equivalent width of \MgII and associated error.
Based on the Kaplan-Meier curves,
we infer a median \ewr\ $\rm \approx 0.39\pm0.08\AA$ for galaxies with 
$\Phi<45\degree$ and \ewr\ $\rm \approx 0.53^{+0.13}_{-0.11}\AA$ for 
galaxies with $\Phi \geq 45\degree$.
The excess \ewr\ around galaxies closer to the minor axis 
($\Phi \geq 45\degree$) is at $\lesssim 1\sigma$ level.
We find no statistical significant dependence of $\left<\kappa\right>$ or 
\ewr\ on azimuthal angle $\Phi$.

\subsection{Comparison between Star-forming and Quiescent Galaxies}
With the large isolated galaxy samples, here we compare 
the physical properties of \MgII absorbing gas around star-forming 
and quiescent galaxies.  Using our likelihood analysis, 
we have noted that there is a significant anti-correlation 
($\sim3\sigma$) between the \MgII absorption strength and 
projected distance $d$ for star-forming galaxies.  
The anti-correlation becomes even stronger when including 
scaling with $B$-band luminosity ($\sim6\sigma$) and 
stellar mass ($\sim 8\sigma$).  In contrast, the results 
of likelihood analysis show that the quiescent galaxies 
do not have a trend in \MgII absorption strength versus 
projected distance within a $\sim1\sigma$ level.  
Including the scaling of $B$-band luminosity or stellar mass 
has little improvement on the anti-correlation.  

In Figure \ref{figure:kappa1}, we show that both star-forming 
and quiescent galaxies show steep declining covering fraction 
$\left<\kappa\right>$ with increasing $d$.  
For the star-forming galaxy sample, the covering fraction of 
${W}_r(2796)\geq 0.3\,$\AA\ absorbers
 declines from $\left<\kappa\right>\approx0.83$ at $d<40\,\rm kpc$ to $\left<\kappa\right>\approx0.37$ at 
 $d\approx40-90\,\rm kpc$ ($\sim$55\% decline), similar to 
 that of the quiescent galaxy sample ($\sim$ 51\% decline).
 Beyond $d\approx100\,\rm kpc$, the covering fraction of both 
 samples is consistent with $\left<\kappa\right>\approx 0$.
 The dependence of $\left<\kappa\right>$ and $d$ for our 
 galaxies are in stark contrast to the massive LRG samples 
 \citep{Huang:2016}.  Specifically, the passive LRGs display 
 flat distribution of $\left<\kappa\right>\,\sim$15\% at 
 $d\lesssim120$ kpc, and an overall $\left<\kappa\right>\,\sim$5\% out to $d\sim500$ kpc.

Next, in Figure \ref{figure:LRGs}(a) we show that in addition 
to the strong dependence of covering fraction on galaxy mass, 
at a similar mass range of $\log M_{\rm star}/\msun\approx{10.4-11.1}$, 
star-forming galaxies reveal an elevated covering fraction than that 
of quiescent galaxies. Note that despite the higher $\left<\kappa\right>$ 
we obtain for the star-forming galaxies, the sample has on average 
a slightly lower stellar mass $\log \langle M_{\rm star}/\msun\rangle$ 
of $\sim 10.5$ compared to that of the quiescent galaxy sample 
($\log \langle M_{\rm star}/\msun\rangle\sim 10.7$).
To investigate whether the elevated $\left<\kappa\right>$ for 
star-forming galaxies is due to  the physical properties of 
associated galaxies or simply a steep dependence on mass, 
we restrict the two samples to a narrow stellar mass range 
of $\log M_{\rm star}/\msun=10.4-10.7$, making both samples 
similar stellar mass distributions with 
$\log \langle M_{\rm star}/\msun\rangle=10.55$.  We find the 
resultant covering fraction of star-forming galaxies is $0.64\pm0.10$, 
while that of quiescent galaxies is $0.29\pm0.13$.  With the same stellar 
mass distribution, the quiescent galaxies have merely 45\% gas 
covering fraction compared to the star-forming galaxies.  
We find the resultant covering fraction of star-forming galaxies 
is $0.64\pm0.10$, while that of quiescent galaxies is $0.29\pm0.13$.  
With the same stellar mass distribution, the quiescent galaxies have 
merely 45\% gas covering fraction compared to the 
star-forming galaxies.  The positive correlation between specific 
star formation rate and gas covering fraction is 
manifest in Figure \ref{figure:LRGs}(b).
An enhanced \MgII covering fraction around star-forming galaxies 
seems to imply an outflow origin.
However, the origin is complicated by the fact that 
the majority of \MgII absorbers around isolated galaxies are 
gravitationally bound (Figure \ref{figure:vdiff_isosf}).  In addition,
no statistically significant dependence of $\langle\kappa\rangle$ 
or \ewr\ on the azimuthal angle is found in Section \ref{AngularDistribution}.
Furthermore, the kinematics discussed in Section \ref{Kinematics} indicates
that while cool clumps around star-forming galaxies can reach the 
center of the halo, the clumps in massive quiescent halos are likely 
to be destroyed during the infall before reaching the LRGs
\citep[e.g.,][]{Gauthier:2011,Huang:2016,Zahedy:2019}.

\subsection{Comparison between Isolated and Non-isolated Systems}
In the $right$ panel of Figure \ref{figure:EW1}, we find that the 
43 non-isolated systems exhibit no hint of a trend between \MgII 
absorber strength \ewr\ and galaxy projected distance $d$ among 
detections, contrary to the clear anti-correlation shown in isolated 
galaxies ($central$ panel). Specifically, while we do not find any 
detection beyond $\sim 100\,$kpc for isolated, star-forming galaxies, 
strong systems of \ewr\ $\gtrsim 0.5$\AA\ are detected for non-isolated 
systems at large distances. 
{The results remain unchanged 
irrespective of whether the distance from nearest galaxy 
or a luminosity-weighted distance is adopted for non-isolated systems.}
The result is in line with previous findings that detections of 
\ewr\ $\gtrsim 0.5$ \AA\ absorbers are frequently found in non-isolated 
or group systems at beyond $d\sim$100 kpc \citep[e.g.,][]{Chen:2010,Nielsen:2018,Fossati:2019}.
Indeed, previous studies have discovered strong MgII absorbers 
\ewr\ $>$1 \AA\ associated with galaxy groups 
\citep[e.g.,][]{Whiting:2006,Fossati:2019}, 
LRGs \citep[][]{Gauthier:2013,Huang:2016} and luminous quasar 
hosts \citep[]{Johnson:2015},
whereas such absorbers are only found at $d\lesssim50$ kpc 
in our 211 isolated galaxy samples.

Recently, new wide-field integral field spectrographs such as 
the Multi Unit Spectroscopic Explorer \citep[MUSE;][]{Bacon:2010}
have enabled discoveries of spatially extended line-emitting 
nebulae on scales reaching $\sim100$ kpc in group or 
cluster environments \citep[e.g.,][]{Epinat:2018,Johnson:2018,Chen:2019}.
\cite{Johnson:2018} found that these giant \OIII\ nebulae
correspond both morphologically and kinematically to 
interacting galaxy pairs in the group,
likely arising from cool filaments and interaction-related debris.
Furthermore, covering $\sim20\%$ of the area around the quasar
at $\lesssim100$ kpc, the nebulae may be an explanation of 
the high covering fraction of \MgII absorbing gas around luminous
QSO hosts in \cite{Johnson:2015}.
\cite{Chen:2019} also uncovered a giant nebula ($\sim100$ physical kpc)
associated with a low-mass galaxy group at $z\sim0.3$, 
where the line-emitting gas connects between group 
galaxies and follows closely the motion of member galaxies.
The study demonstrates that gas stripping in low-mass groups 
may be effective in releasing metal-enriched gas from
star-forming regions, producing absorption systems 
(e.g., \MgII absorbers) in QSO spectra.

The case studies of spatially-extended line-emitting nebulae 
associated with galaxy groups provide unambiguous evidence 
of the importance of interactions in distributing metal-enriched 
gas on large scales. Despite the lack of morphological 
information in our absorption-line survey, 
the commonly found strong \MgII absorbers 
(\ewr\ $\gtrsim 0.5$ \AA) at $d\gtrsim100$ kpc in the 43 non-isolated 
systems support the idea that interactions between group galaxies 
may contribute to the presence of strong absorbers at large scales.

\subsection{Comparison with Other Studies}
The \MAGMAGMAG\ Halo Project consists of 211 isolated 
and 43 non-isolated galaxies with $z=0.10-0.48$ at projected 
distance $\langle d \rangle_{\rm med}=73$ kpc from a 
background QSO, chosen without any prior knowledge 
of the presence or absence of \MgII absorbing gas.
The absorption-blind sample chosen from SDSS enables
an unbiased characterization of the correlation between 
\MgII absorbing gas and physical properties of associated galaxies.
Our survey shows a distinct difference in the $W_r(2796)$ 
versus $d$  inverse correlation between
star-forming and quiescent halos.  While there is a significant 
anti-correlation ($\gtrsim 3\sigma$) between $W_r(2796)$ 
and $d$ for star-forming galaxies, there is no hint of a correlation 
among \MgII detected quiescent galaxies.
We also show that while star-forming galaxies have elevated 
$\langle \kappa \rangle \approx 83$\% at $d<40$ kpc, 
both star-forming and quiescent galaxies show 
$\langle \kappa \rangle \approx 0$ beyond 90 kpc.
These findings are different from the results of MAG{\scriptsize II}CAT \citep{Nielsen:2013a,Nielsen:2013b}.
 At the $W_{0}=0.3\rm \,\AA$ threshold, MAG{\scriptsize II}CAT 
 galaxies reveal a non-negligible \MgII gas covering fraction 
 $\langle \kappa \rangle \approx 10-40$\% at $d=100-200$ kpc.

It is worth noting that even though 69 out of 182 galaxies 
in MAG{\scriptsize II}CAT come from our \MAGMAGMAG\ Halo Project \citep{Chen:2008,Chen:2010}, MAG{\scriptsize II}CAT consists 
of galaxy-\MgII absorber pairs from different programs.  
While some of these programs were designed to be absorption-blind 
like the \MAGMAGMAG\ Halo Project, others select galaxies at the 
redshifts of known \MgII absorbers.  The galaxy-absorber pairs from 
these other programs have associated \MgII absorbers by design, 
therefore imposing a strong bias on the calculated gas covering fraction.
Similarly, the lack of mass dependence on the covering fraction of 
\MgII absorbing gas in \cite{Churchill:2013} can be understood by an 
over estimation of covering fraction at high mass, where detections 
at this mass range come mostly from absorber centric surveys.

\cite{Martin:2019} constructed a sample of 50 $z\approx0.2$ 
star-forming galaxies with $M_{\rm star}\approx 10^{10}\, M_\odot$ 
at close projected distances $d<100$ kpc, properties similar to our 
star-forming galaxies. For absorbers with a detection threshold of 
$W_0=0.3 \, \rm \AA$, this sample yields 
$\langle \kappa \rangle \approx 75$\% at $d<40$ kpc, which declines 
to {$\langle \kappa \rangle \approx 24$\% }at $40<d<100$ kpc 
(through private communication).  The steep anti-correlation and 
the mean covering fraction is roughly consistent with our results, 
where we find $\left<\kappa\right>$ declines from 
$\left<\kappa\right>=0.83^{+0.06}_{-0.05}$ at $d<40\,\rm kpc$ to $\left<\kappa\right>=0.37^{+0.07}_{-0.07}$ at 
 $d\approx40-90\,\rm kpc$. 
 
It is clear from our results that \MgII absorbing gas is tightly 
coupled with the physical properties of host galaxies.  
In particular, our survey displays a strong dependence 
of mean covering fraction of \MgII absorbing gas 
$\langle \kappa \rangle$ on the stellar mass of host galaxies, 
where $\langle \kappa \rangle$ increases with mass at 
$\log M_{\rm star}/\msun\lesssim10.6$ and decreases 
steeply at higher masses. We further demonstrate in 
Section \ref{Kinematics} that at the same stellar masses 
of $\log \langle M_{\rm star}/\msun\rangle=10.55$, 
star-forming galaxies have in average twice of the 
mean \MgII gas covering fraction (64\%) compared to 
that of quiescent galaxies (29\%).  
The properties of \MgII absorbing gas are shown to strongly 
depend on multiple host galaxy properties. 
Here we highlight the importance of having a homogeneous, 
absorption-blind galaxy sample, in order to identify different 
dependences (e.g. stellar mass and star-formation) and 
carefully study how \MgII gas correlates with their host galaxy properties.

\section{Summary}
\label{section:summary}
We have carried out the \MAGMAGMAG\ Halo Project 
of galaxies and \MgII absorbers in the spectra of 
background QSOs that are within close projected 
distances at $z<0.5$.  The catalog contains 211 isolated 
and 43 non-isolated galaxy-QSO pairs with spectroscopic
redshifts of $\langle z \rangle_{\rm med}=0.21$ and projected 
distances of $\langle d \rangle_{\rm med}=86$ kpc.  
This is the largest homogeneous, 
absorption-blind sample at $z\sim0.2$ to date, allowing us
to conduct a comprehensive study of the correlation 
between \MgII absorbing gas and the physical properties 
of host galaxies at low-redshift. The main findings of our 
survey are summarized as the following:
\begin{enumerate}[leftmargin=.2in]
\item [(1)]We observe a stark contrast in the distribution 
of \MgII absorber strength \ewr\ versus galaxy projected 
distance $d$ between isolated and non-isolated galaxies (Figure \ref{figure:EW1}).  
While both galaxy samples appear to occupy a similar 
\ewr\ versus $d$ space,
isolated galaxies show strong inverse correlation 
but non-isolated galaxies exhibit no hint of a trend among detections.  
When dividing isolated galaxies into star-forming and quiescent 
galaxy samples, star-forming galaxies show a strong anti-correlation 
between \ewr\ and distance $d$, in contrast to the moderate 
trend revealed in quiescent galaxy sample.\

\item[(2)]
Based on the likelihood analysis, we confirm that \ewr\ declines 
with increasing $d$
for isolated galaxies.  The  anti-correlation between \ewr\ and 
$d$ is strengthened when considering only isolated, 
star-forming galaxies.  The inverse correlation is further enhanced 
for star-forming galaxies after accounting for either mass scaling 
of gaseous radius ($R_{\rm h}$), $B$-band luminosity($L_{\rm B}$) 
or stellar mass ($M_{\rm star}$) of host galaxies.
On the contrary, \MgII detected quiescent galaxies exhibit little 
correlation between \ewr\ and $d$, whether or not accounting 
for scaling (see Figures \ref{figure:fit_d} \& \ref{figure:fit_MB}).

\item[(3)]
In Figure \ref{figure:kappa1}, we show that the covering fraction 
of \MgII absorbing gas $\langle \kappa \rangle$ is high for isolated 
galaxies at small projected distances $d$ and declines rapidly to 
$\langle \kappa \rangle \approx 0$ at $d\gtrsim100$ kpc for 
absorbers of  \ewr$\geq0.3\,\rm \AA$.  At $d<40$ kpc, 
we find an elevated covering fraction $\langle \kappa \rangle \sim 0.83$ 
for star-forming galaxies compared to $\langle \kappa \rangle \sim 0.57$ 
for quiescent galaxies.  After the scaling of $B-$band luminosity, 
the inverse correlation between $\langle \kappa \rangle$ and 
$d$ is strengthened and the difference in covering fraction 
between star-forming and quiescent galaxies become 
more evident ($\sim 3\sigma$).

\item[(4)]
The high \MgII gas covering fraction for both our star-forming 
and quiescent galaxy samples at $d<90\,$kpc 
($\langle \kappa \rangle \sim 0.3-0.8$) is in stark contrast 
to what we have seen in the massive LRGs.  
Within $R_{\rm gas}$, we find a sharp decline of 
\MgII covering fraction $\langle \kappa \rangle$ from 
30-70\% around $L_*$ galaxies to 10-15\% around 
massive LRGs (Figure \ref{figure:LRGs}a).  
The strong mass dependence of \MgII incidence is qualitatively 
consistent with the expectation from the 
observed clustering of \MgII absorbers.
In addition, at stellar mass of $\log \langle M_{\rm star}/\msun\rangle \approx{10.6}$, the \MgII gas covering fraction for quiescent galaxies 
($\langle \kappa \rangle=0.29$) is merely half of what we find 
for star-forming galaxies ($\langle \kappa \rangle$=0.64).
We also find a positive correlation between specific star 
formation rate and \MgII gas covering fraction (Figure \ref{figure:LRGs}b).

\item[(5)]
We find that most of the galaxy-\MgII absorber pairs have 
relative velocities smaller than the expected projected 
escape velocity of their host halos, implying that
the \MgII absorbers are likely to be gravitationally bound 
(Figure \ref{figure:vdiff_isosf}). In addition, \MgII absorbers 
have line-of-sight velocity dispersion of  $\sigma$= (58-73, 73-90) \kms 
for low-mass and high-mass star-forming galaxies, consistent with the 
expected line-of-sight velocity dispersion $\sigma$= (52, 86) \kms 
for virialized motion. If the clouds are massive enough to travel 
through the hot gas at the halo velocity without
significant deceleration by the hot gas drag force, we are able 
to place lower limits on the cloud mass of 
$m_{\rm cl} \sim 10^5 \,\rm M_\odot$ and the cool gas accretion 
rate of $\sim 0.7-2 \,M_\odot\,\rm yr^{-1}$ .

\item[(6)] 
In Figure \ref{figure:PA},
we investigate the possible azimuthal dependence 
in the covering fraction of \MgII absorbers for isolated, 
star-forming and quiescent galaxies.  While no apparent trend 
is seen for quiescent galaxies at $d\leq0.4\,R_{\rm h}$,
there is a modest enhancement in the gas covering fraction 
along the minor axis
of star-forming galaxies at $0.2<d\leq0.4\,R_{\rm h}$.
We find excess \ewr\ around galaxies closer to the minor
axis ($\Phi \leq 45 \degree$) at $\lesssim 1 \sigma$ level.
No statistical significant dependence of 
$\left<\kappa\right>_{\rm MgII}$ or 
\ewr\ on azimuthal angle $\Phi$ is shown in our isolated galaxy sample.

\end{enumerate}

\section*{Acknowledgments}
The authors thank Stephanie Ho for the covering 
fraction measurement for their survey and careful reading 
on the early draft of this paper.
HWC acknowledges partial support from HST-GO- 15163.001A and NSF AST-1715692 grants.
YHH acknowledges support from NSF AST 15-15115 and AST 19-08284 grants.

We are grateful to the SDSS collaboration for producing and maintaining the SDSS public data archive.  Funding for the SDSS and SDSS-II has been provided by the Alfred P. Sloan Foundation, the Participating Institutions, the National Science Foundation, the U.S. Department of Energy, the National Aeronautics and Space Administration, the Japanese Monbukagakusho, the Max Planck Society, and the Higher Education Funding Council for England. The SDSS Web Site is http://www.sdss.org/.  The SDSS is managed by the Astrophysical Research Consortium for the Participating Institutions.

This research has made use of NASA?s Astrophysics Data System and the NASA/IPAC Extragalactic Database (NED) which is operated by the Jet Propulsion Laboratory, California Institute of Technology, under con- tract with the National Aeronautics and Space Administration.

\section*{DATA AVAILABILITY}
The data underlying this article will be shared on reasonable request to the corresponding author.

\footnotesize{
\bibliographystyle{mnras} 
\bibliography{biblio}
}

\bsp

\label{lastpage}


\clearpage




\end{document}